\pdfoutput=1

\documentclass[12pt,a4paper]{article}

\usepackage{ifthen} 
\newboolean{pdflatex}
\setboolean{pdflatex}{true} 

\newboolean{articletitles}
\setboolean{articletitles}{true} 

\newboolean{uprightparticles}
\setboolean{uprightparticles}{false} 

\usepackage{microtype}
\usepackage{lineno}  % for line numbering during review
\usepackage{xspace} % To avoid problems with missing or double spaces after
\usepackage{graphicx}  % to include figures (can also use other packages)
\usepackage{color}
\usepackage{colortbl}
\graphicspath{{./figs/}} % Make Latex search fig subdir for figures

\usepackage{amsmath} % Adds a large collection of math symbols
\usepackage{amssymb}
\usepackage{amsfonts}
\usepackage{upgreek} % Adds in support for greek letters in roman typeset

\newcommand*\patchAmsMathEnvironmentForLineno[1]{%
\expandafter\let\csname old#1\expandafter\endcsname\csname #1\endcsname
\expandafter\let\csname oldend#1\expandafter\endcsname\csname
end#1\endcsname
 \renewenvironment{#1}%
   {\linenomath\csname old#1\endcsname}%
   {\csname oldend#1\endcsname\endlinenomath}%
}
\newcommand*\patchBothAmsMathEnvironmentsForLineno[1]{%
  \patchAmsMathEnvironmentForLineno{#1}%
  \patchAmsMathEnvironmentForLineno{#1*}%
}
\AtBeginDocument{%
\patchBothAmsMathEnvironmentsForLineno{equation}%
\patchBothAmsMathEnvironmentsForLineno{align}%
\patchBothAmsMathEnvironmentsForLineno{flalign}%
\patchBothAmsMathEnvironmentsForLineno{alignat}%
\patchBothAmsMathEnvironmentsForLineno{gather}%
\patchBothAmsMathEnvironmentsForLineno{multline}%
}

\usepackage{hyperref}    % Hyperlinks in references
\usepackage[all]{hypcap} % Internal hyperlinks to floats.

\def\lhcb {\mbox{LHCb}\xspace}
\def\ux85 {\mbox{UX85}\xspace}

\def\babar  {\mbox{BaBar}\xspace}

\ifthenelse{\boolean{uprightparticles}}%
{

 \def\Pmu         {\ensuremath{\upmu}\xspace}

 \def\Ppi         {\ensuremath{\uppi}\xspace}

 \def\Ppsi        {\ensuremath{\uppsi}\xspace}

 \def\PDelta      {\ensuremath{\Delta}\xspace}                 
 \def\PXi      {\ensuremath{\Xi}\xspace}                 
 \def\PLambda      {\ensuremath{\Lambda}\xspace}                 
 \def\PSigma      {\ensuremath{\Sigma}\xspace}                 
 \def\POmega      {\ensuremath{\Omega}\xspace}                 
 \def\PUpsilon      {\ensuremath{\Upsilon}\xspace}                 
 
 \def\PB      {\ensuremath{\mathrm{B}}\xspace}                 
                  
 \def\PD      {\ensuremath{\mathrm{D}}\xspace}

 \def\PJ      {\ensuremath{\mathrm{J}}\xspace}                 
 \def\PK      {\ensuremath{\mathrm{K}}\xspace}

 \def\Pb      {\ensuremath{\mathrm{b}}\xspace}                 
 \def\Pc      {\ensuremath{\mathrm{c}}\xspace}

 \def\Pi      {\ensuremath{\mathrm{i}}\xspace}

 \def\Ps      {\ensuremath{\mathrm{s}}\xspace}

}
{

 \def\Pmu         {\ensuremath{\mu}\xspace}

 \def\Ppi         {\ensuremath{\pi}\xspace}

 \def\Ppsi        {\ensuremath{\psi}\xspace}                 
                  
 \mathchardef\PDelta="7101
 \mathchardef\PXi="7104
 \mathchardef\PLambda="7103
 \mathchardef\PSigma="7106
 \mathchardef\POmega="710A
 \mathchardef\PUpsilon="7107
                  
 \def\PB      {\ensuremath{B}\xspace}                 
                  
 \def\PD      {\ensuremath{D}\xspace}

 \def\PJ      {\ensuremath{J}\xspace}                 
 \def\PK      {\ensuremath{K}\xspace}

 \def\Pb      {\ensuremath{b}\xspace}                 
 \def\Pc      {\ensuremath{c}\xspace}

 \def\Pi      {\ensuremath{i}\xspace}

 \def\Ps      {\ensuremath{s}\xspace}

}

\def\mup        {\ensuremath{\Pmu^+}\xspace}
\def\mun        {\ensuremath{\Pmu^-}\xspace} % muon negative (\mum is taken)
\def\mumu       {\ensuremath{\Pmu^+\Pmu^-}\xspace}

\def\squark    {\ensuremath{\Ps}\xspace}

\def\cquark    {\ensuremath{\Pc}\xspace}
\def\cquarkbar {\ensuremath{\overline \cquark}\xspace}
\def\ccbar     {\ensuremath{\cquark\cquarkbar}\xspace}
\def\bquark    {\ensuremath{\Pb}\xspace}

\def\pion  {\ensuremath{\Ppi}\xspace}

\def\pip   {\ensuremath{\pion^+}\xspace}
\def\pim   {\ensuremath{\pion^-}\xspace}

\def\kaon  {\ensuremath{\PK}\xspace}
  \def\Kbar  {\kern 0.2em\overline{\kern -0.2em \PK}{}\xspace}

\def\Kz    {\ensuremath{\kaon^0}\xspace}
\def\Kzb   {\ensuremath{\Kbar^0}\xspace}
\def\KzKzb {\ensuremath{\Kz \kern -0.16em \Kzb}\xspace}
\def\Kp    {\ensuremath{\kaon^+}\xspace}
\def\Km    {\ensuremath{\kaon^-}\xspace}

\def\KpKm  {\ensuremath{\Kp \kern -0.16em \Km}\xspace}

\def\Kstarz  {\ensuremath{\kaon^{*0}}\xspace}
\def\Kstarzb {\ensuremath{\Kbar^{*0}}\xspace}

  \def\Dbar    {\kern 0.2em\overline{\kern -0.2em \PD}{}\xspace}
\def\D       {\ensuremath{\PD}\xspace}

\def\Dz      {\ensuremath{\D^0}\xspace}
\def\Dzb     {\ensuremath{\Dbar^0}\xspace}
\def\DzDzb   {\ensuremath{\Dz {\kern -0.16em \Dzb}}\xspace}
\def\Dp      {\ensuremath{\D^+}\xspace}
\def\Dm      {\ensuremath{\D^-}\xspace}

\def\DpDm    {\ensuremath{\Dp {\kern -0.16em \Dm}}\xspace}

\def\B       {\ensuremath{\PB}\xspace}
  \def\Bbar    {\kern 0.18em\overline{\kern -0.18em \PB}{}\xspace}

\def\Bz      {\ensuremath{\B^0}\xspace}
\def\Bzb     {\ensuremath{\Bbar^0}\xspace}
\def\Bu      {\ensuremath{\B^+}\xspace}

\def\Bp      {\ensuremath{\Bu}\xspace}

\def\Bs      {\ensuremath{\B^0_\squark}\xspace}
\def\Bsb     {\ensuremath{\Bbar^0_\squark}\xspace}

\def\jpsi     {\ensuremath{{\PJ\mskip -3mu/\mskip -2mu\Ppsi\mskip 2mu}}\xspace}
\def\psitwos  {\ensuremath{\Ppsi{(2S)}}\xspace}

  \def\Y#1S{\ensuremath{\PUpsilon{(#1S)}}\xspace}% no space before {...}!

\def\L {\ensuremath{\PLambda}\xspace}
\def\Lbar {\ensuremath{\kern 0.1em\overline{\kern -0.1em\PLambda}}\xspace}

\def\Lb      {\ensuremath{\L^0_\bquark}\xspace}

\def\BF         {{\ensuremath{\cal B}\xspace}}

\newcommand{\decay}[2]{\ensuremath{#1\!\to #2}\xspace}         % {\Pa}{\Pb \Pc}

\def\to                 {\ensuremath{\rightarrow}\xspace}

\def\qsq       {\ensuremath{q^2}\xspace}

\def\CP                {\ensuremath{C\!P}\xspace}

\def\AT#1     {\ensuremath{A_{\mathrm{T}}^{#1}}\xspace}           % 2

\def\C#1      {\ensuremath{\mathcal{C}_{#1}}\xspace}                       % 9
\def\Cp#1     {\ensuremath{\mathcal{C}_{#1}^{'}}\xspace}                    % 7
\def\Ceff#1   {\ensuremath{\mathcal{C}_{#1}^{\mathrm{(eff)}}}\xspace}        % 9  
\def\Cpeff#1  {\ensuremath{\mathcal{C}_{#1}^{'\mathrm{(eff)}}}\xspace}       % 7
\def\Ope#1    {\ensuremath{\mathcal{O}_{#1}}\xspace}                       % 2
\def\Opep#1   {\ensuremath{\mathcal{O}_{#1}^{'}}\xspace}                    % 7

\newcommand{\tev}{\ensuremath{\mathrm{\,Te\kern -0.1em V}}\xspace}
\newcommand{\gev}{\ensuremath{\mathrm{\,Ge\kern -0.1em V}}\xspace}
\newcommand{\mev}{\ensuremath{\mathrm{\,Me\kern -0.1em V}}\xspace}
\newcommand{\kev}{\ensuremath{\mathrm{\,ke\kern -0.1em V}}\xspace}
\newcommand{\ev}{\ensuremath{\mathrm{\,e\kern -0.1em V}}\xspace}
\newcommand{\gevc}{\ensuremath{{\mathrm{\,Ge\kern -0.1em V\!/}c}}\xspace}
\newcommand{\mevc}{\ensuremath{{\mathrm{\,Me\kern -0.1em V\!/}c}}\xspace}
\newcommand{\gevcc}{\ensuremath{{\mathrm{\,Ge\kern -0.1em V\!/}c^2}}\xspace}
\newcommand{\gevgevcccc}{\ensuremath{{\mathrm{\,Ge\kern -0.1em V^2\!/}c^4}}\xspace}
\newcommand{\mevcc}{\ensuremath{{\mathrm{\,Me\kern -0.1em V\!/}c^2}}\xspace}

\def\mum  {\ensuremath{\,\upmu\rm m}\xspace}

\def\invpb {\ensuremath{\mbox{\,pb}^{-1}}\xspace}

\def\invfb   {\ensuremath{\mbox{\,fb}^{-1}}\xspace}

\def\deriv {\ensuremath{\mathrm{d}}}

\def\gsim{{~\raise.15em\hbox{$>$}\kern-.85em
          \lower.35em\hbox{$\sim$}~}\xspace}
\def\lsim{{~\raise.15em\hbox{$<$}\kern-.85em
          \lower.35em\hbox{$\sim$}~}\xspace}

\def\pt         {\mbox{$p_{\rm T}$}\xspace}

\def\evtgen     {\mbox{\textsc{EvtGen}}\xspace}
\def\pythia     {\mbox{\textsc{Pythia}}\xspace}

\def\geant      {\mbox{\textsc{Geant4}}\xspace}

\def\photos     {\mbox{\textsc{Photos}}\xspace}

\def\tell1  {TELL1\xspace}
\def\ukl1   {UKL1\xspace}

\usepackage{cite} % Allows for ranges in citations
\usepackage{mciteplus}

\begin{document}

\renewcommand{\thefootnote}{\fnsymbol{footnote}}
\setcounter{footnote}{1}

\hypersetup{pageanchor=false}

\begin{titlepage}
\pagenumbering{roman}

\vspace*{-1.5cm}
\centerline{\large EUROPEAN ORGANIZATION FOR NUCLEAR RESEARCH (CERN)}
\vspace*{1.5cm}
\hspace*{-0.5cm}
\begin{tabular*}{\linewidth}{lc@{\extracolsep{\fill}}r}
\ifthenelse{\boolean{pdflatex}}
{\vspace*{-2.7cm}\mbox{\!\!\!\includegraphics[width=.14\textwidth]{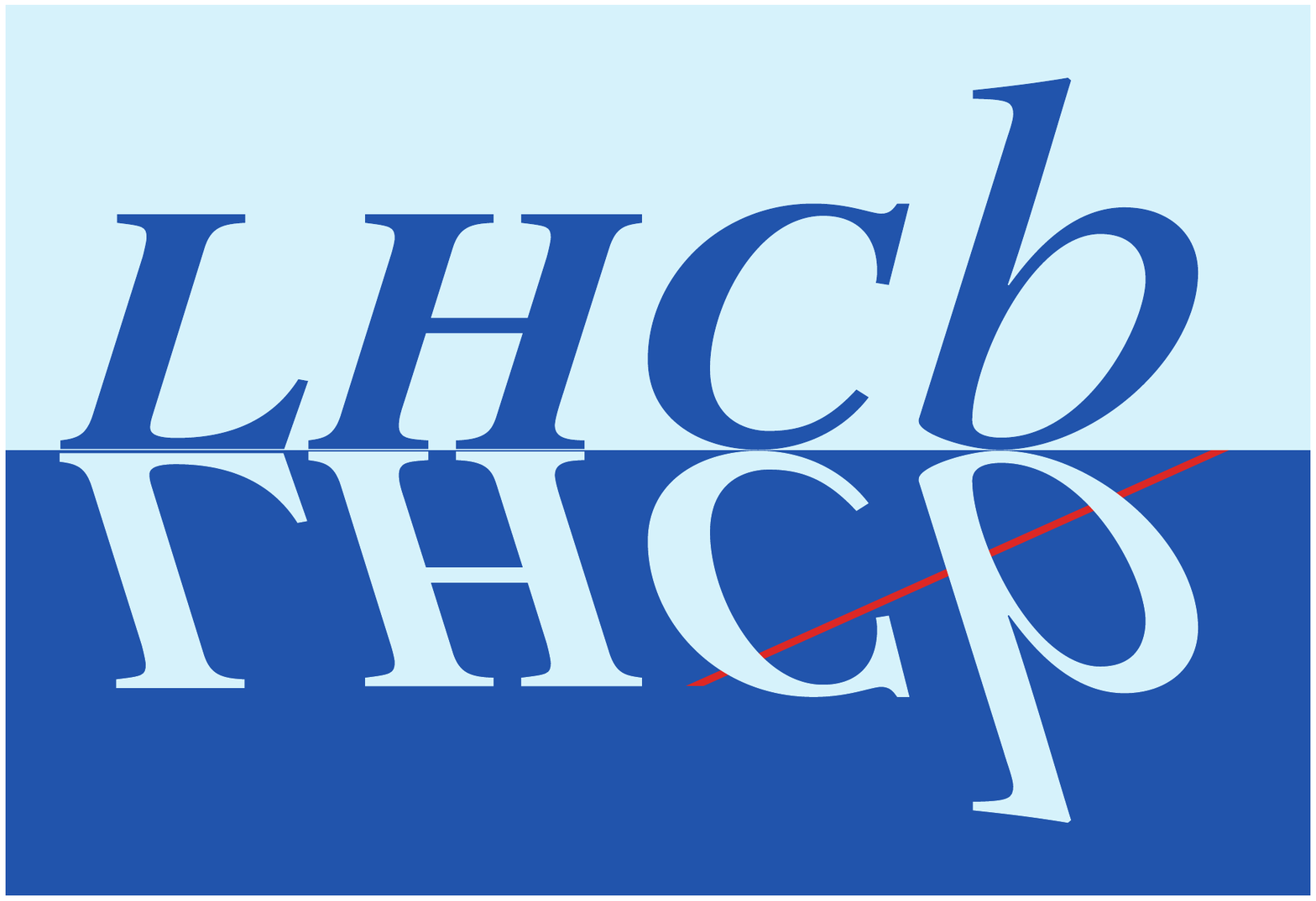}} & &}%
{\vspace*{-1.2cm}\mbox{\!\!\!\includegraphics[width=.12\textwidth]{figs/lhcb-logo.eps}} & &}%
\\
 & & CERN-PH-EP-2013-074 \\  % ID 
 & & LHCb-PAPER-2013-019 \\  % ID 
 & & 8 July 2013 \\ 
 & &  \\
\end{tabular*}

\vspace*{3.0cm}

{\bf\boldmath\huge
\begin{center}
  Differential branching fraction \\ and angular analysis of \\ the decay \decay{\Bz}{\Kstarz\mumu}
\end{center}
}

\vspace*{1.8cm}

\begin{center}
The LHCb collaboration\footnote{Authors are listed on the following pages.}
\end{center}

\vspace{\fill}

\begin{abstract}
  \noindent   The angular distribution and differential branching fraction of the decay \mbox{\decay{\Bz}{\Kstarz\mumu}} are studied using a data sample, collected by the \lhcb experiment in $pp$ collisions at $\sqrt{s}=7\tev$, corresponding to an integrated luminosity of $1.0\invfb$. Several angular observables are measured in bins of the dimuon invariant mass squared, \qsq. A first measurement of the zero-crossing point of the forward-backward asymmetry of the dimuon system is also presented. The zero-crossing point is measured to be $q_{0}^{2} = 4.9 \pm 0.9 \gev^{2}/c^{4}$, where the uncertainty is the sum of statistical and systematic uncertainties. The results are consistent with the Standard Model predictions.
\end{abstract}

\vspace*{1.8cm}

\begin{center}
Submitted to JHEP
\end{center}

\vspace{\fill}

{\footnotesize 
\centerline{\copyright~CERN on behalf of the \lhcb collaboration, license \href{http://creativecommons.org/licenses/by/3.0/}{CC-BY-3.0}.}}
\vspace*{2mm}

\end{titlepage}

\newpage
\setcounter{page}{2}
\mbox{~}
\newpage

%%%%%%%%%%%%%%%%%%%%%%%%%%%%%%%%%%%%%%%%%%
\centerline{\large\bf LHCb collaboration}
\begin{flushleft}
\small
R.~Aaij$^{40}$, 
C.~Abellan~Beteta$^{35,n}$, 
B.~Adeva$^{36}$, 
M.~Adinolfi$^{45}$, 
C.~Adrover$^{6}$, 
A.~Affolder$^{51}$, 
Z.~Ajaltouni$^{5}$, 
J.~Albrecht$^{9}$, 
F.~Alessio$^{37}$, 
M.~Alexander$^{50}$
% ...

{\footnotesize \it
% ...
$ ^{5}$Clermont Universit\'{e}, Universit\'{e} Blaise Pascal, CNRS/IN2P3, LPC, Clermont-Ferrand, France\\
$ ^{6}$CPPM, Aix-Marseille Universit\'{e}, CNRS/IN2P3, Marseille, France\\
% ...
$ ^{9}$Fakult\"{a}t Physik, Technische Universit\"{a}t Dortmund, Dortmund, Germany\\
% ...
$ ^{35}$Universitat de Barcelona, Barcelona, Spain\\
$ ^{36}$Universidad de Santiago de Compostela, Santiago de Compostela, Spain\\
$ ^{37}$European Organization for Nuclear Research (CERN), Geneva, Switzerland\\
% ...
$ ^{40}$Nikhef National Institute for Subatomic Physics, Amsterdam, The Netherlands\\
% ...
$ ^{45}$H.H. Wills Physics Laboratory, University of Bristol, Bristol, United Kingdom\\
% ...
$ ^{50}$School of Physics and Astronomy, University of Glasgow, Glasgow, United Kingdom\\
$ ^{51}$Oliver Lodge Laboratory, University of Liverpool, Liverpool, United Kingdom\\
% ...
$ ^{n}$LIFAELS, La Salle, Universitat Ramon Llull, Barcelona, Spain\\
% ...
$ ^{s}$Scuola Normale Superiore, Pisa, Italy\\
}
\end{flushleft}
%%%%%%%%%%%%%%%%%%%%%%%%%%%%%%%%%%%%%%%%%%

\cleardoublepage

\renewcommand{\thefootnote}{\arabic{footnote}}
\setcounter{footnote}{0}

\pagestyle{plain} 
\setcounter{page}{1}
\pagenumbering{arabic}

%\linenumbers

\cleardoublepage

\hypersetup{pageanchor=true}

\section{Introduction}
\label{sec:introduction}

The \decay{\Bz}{\Kstarz\mumu} decay,\footnote{Charge conjugation is implied throughout this paper unless stated otherwise.} where \decay{\Kstarz}{\Kp\pim}, is a $b \to s$ flavour changing neutral current process that is mediated by electroweak box and penguin type diagrams  in the Standard Model (SM). The angular distribution of the $\Kp\pim\mumu$ system offers particular sensitivity to contributions from new particles in extensions to the SM. The differential branching fraction of the decay also provides information on the contribution from those new particles but typically suffers from larger theoretical uncertainties due to hadronic form factors.

The angular distribution of the decay can be described by three angles ($\theta_{\ell},\theta_{K}$ and $\phi$) and by the invariant mass squared of the dimuon system (\qsq). The \decay{\Bz}{\Kstarz\mumu} decay is self-tagging through the charge of the kaon and so there is some freedom in the choice of the angular basis that is used to describe the decay. In this paper, the angle $\theta_{\ell}$ is defined as the angle between the direction of the \mup (\mun) and the direction opposite that of the \Bz (\Bzb) in the dimuon rest frame. The angle $\theta_{K}$ is defined as the angle between the direction of the kaon and the direction of opposite that of the \Bz (\Bzb) in in the \Kstarz (\Kstarzb) rest frame. The angle $\phi$ is the angle between the plane containing the \mup and \mun and the plane containing the kaon and pion from the \Kstarz (\Kstarzb) in the \Bz (\Bzb) rest frame.  The basis is designed such that the angular definition for the \Bzb decay is a \CP transformation of that for the \Bz decay.  This basis differs from some that appear in the literature. A graphical representation, and a more detailed description, of the angular basis is given in Appendix~\ref{sec:appendix:basis}. 

Using the notation of Ref.~\cite{Altmannshofer:2008dz}, the decay distribution of the \Bz corresponds to

{
\begin{equation}
\begin{split}
\frac{\deriv^{4}\Gamma}{\deriv\qsq\,\deriv\cos\theta_{\ell}\,\deriv\cos\theta_{K}\,\deriv\phi} = \frac{9}{32\pi} & 
\left[ \frac{}{} {I_{1}^{s}} \sin^{2} \theta_{K} + 
{I_{1}^{c}}  \cos^{2}\theta_{K}  ~+ \right.  \\ 
&   \left. ~\frac{}{}   {I_{2}^{s}} \sin^{2} \theta_{K} \cos 2\theta_{\ell}  + 
{I_{2}^{c}} \cos^{2} \theta_{K} \cos 2\theta_{\ell} ~ + \right. \\ 
&   \left. ~\frac{}{} {I_{3}}  \sin^{2}\theta_{K} \sin^{2} \theta_{\ell} \cos 2\phi  + {{ I_{4} \sin 2\theta_{K} \sin 2\theta_{\ell} \cos\phi}} ~+ \right. \\ 
& ~\frac{}{} \left. {{{I_{5}} \sin 2\theta_{K} \sin\theta_{\ell}\cos\phi}} + I_{6} \sin^{2}\theta_{K}  \cos\theta_{\ell} ~+ \right. \\ 
& ~\frac{}{} \left. {{{I_{7}}  \sin 2\theta_{K} \sin\theta_{\ell} \sin\phi}} +  
{{{I_{8}} \sin 2\theta_{K} \sin 2\theta_{\ell}\sin\phi}} ~+ \right. \\ 
& ~\frac{}{} \left. I_{9} \sin^{2} \theta_{K} \sin^{2}\theta_{\ell} \sin 2\phi \frac{}{}~\right] ~,
\end{split}
\label{eq:fullangular}
\end{equation}
}

\noindent where the 11 coefficients, $I_{j}$, are bilinear combinations of \Kstarz decay amplitudes, ${\cal A}_{m}$, and vary with \qsq. The superscripts $s$ and $c$ in the first two terms arise in Ref.~\cite{Altmannshofer:2008dz} and indicate either a $\sin^{2}\theta_{K}$ or $\cos^{2}\theta_{K}$ dependence of the corresponding angular term. In the SM, there are seven complex decay amplitudes, corresponding to different polarisation states of the \Kstarz and chiralities of the dimuon system. In the angular coefficients, the decay amplitudes appear in the combinations $|{\cal A}_{m}|^{2}$, ${\rm Re}({\cal A}_{m}{\cal A}_{n}^{*})$ and ${\rm Im}({\cal A}_{m}{\cal A}_{n}^{*})$. Combining \Bz and \Bzb decays, and assuming there are equal numbers of each, it is possible to build angular observables that depend on the average of, or difference between, the distributions for the \Bz and \Bzb decay, 

\begin{equation}
S_{j} = \left. \left( I_{j} + \bar{I}_{j} \right) \middle/ \frac{\deriv\Gamma}{\deriv\qsq}  \right. ~\text{or}~ \left. A_{j} = \left( I_{j} - \bar{I}_{j} \right) \middle/ \frac{\deriv\Gamma}{\deriv\qsq} \right. ~.
\end{equation} 

These observables are referred to below as \CP averages or \CP asymmetries and are normalised with respect to the combined differential decay rate, $\deriv\Gamma/\deriv\qsq$, of \Bz and \Bzb decays.  The observables $S_7$, $S_8$ and $S_9$ depend on combinations ${\rm Im}({\cal A}_{m}{\cal A}_{n}^{*})$ and are suppressed by the small size of the strong phase difference between the decay amplitudes. They are consequently expected to be close to zero across the full \qsq range not only in the SM but also in most extensions. However, the corresponding \CP asymmetries, $A_7$, $A_8$ and $A_9$, are not suppressed by the strong phases involved~\cite{Bobeth:2008ij} and remain sensitive to the effects of new particles. 

If the \Bz and \Bzb decays are combined using the angular basis in Appendix~\ref{sec:appendix:basis}, the resulting angular distribution is sensitive to only the \CP averages of each of the angular terms. Sensitivity to $A_7$, $A_8$ and $A_9$ is achieved by flipping the sign of $\phi$ ($\phi \to -\phi$) for the \Bzb decay. This procedure results in a combined \Bz and \Bzb angular distribution that is sensitive to the \CP averages $S_{1} - S_{6}$ and the \CP asymmetries of $A_7$, $A_8$ and $A_9$. 

In the limit that the dimuon mass is large compared to the mass of the muons, $\qsq \gg 4m_{\mu}^{2}$, the \CP average of $I_1^c$, $I_1^s$, $I_2^c$ and $I_2^s$ ($S_1^c$, $S_1^s$, $S_2^c$ and $S_2^s$) are related to the fraction of longitudinal polarisation of the \Kstarz meson, $F_{\rm L}$ ($S_1^{c} = -S_2^{c} = F_{\rm L}$ and $\frac{4}{3} S_1^{s} = 4 S_2^{s} = 1-F_{\rm L}$). The angular term, $I_{6}$ in Eq.~\ref{eq:fullangular}, which has a $\sin^{2}\theta_{K} \cos\theta_{\ell}$ dependence, generates a forward-backward asymmetry of the dimuon system, $A_{\rm FB}$~\cite{Ali:1999mm} ($A_{\rm FB} = \frac{3}{4}S_6$). The term $S_{3}$ is related to the asymmetry between the two sets of transverse \Kstarz amplitudes, referred to in literature as $A_{\rm T}^{2}$~\cite{Kruger:2005ep}, where $S_3 = \frac{1}{2}\left( 1-F_{\rm L} \right) A_{\rm T}^{2}$. 

In the SM, $A_{\rm FB}$ varies as a function of \qsq and is known to change sign. The \qsq dependence arises from the interplay between the different penguin and box diagrams that contribute to the decay. The position of the zero-crossing point of $A_{\rm FB}$ is a precision test of the SM since, in the limit of large \Kstarz energy, its prediction is free from form-factor uncertainties~\cite{Ali:1999mm}. At large recoil, low values of \qsq, penguin diagrams involving a virtual photon dominate. In this \qsq region, $A_{\rm T}^{2}$ is sensitive to the polarisation of the virtual photon which, in the SM, is predominately left-handed, due to the nature of the charged-current interaction. In many possible extensions of the SM however, the photon can be both left- or right-hand polarised, leading to large enhancements of $A_{\rm T}^{2}$~\cite{Kruger:2005ep}.

The one-dimensional $\cos\theta_\ell$ and $\cos\theta_K$ distributions have previously been studied by the \lhcb~\cite{Aaij:2011aa}, \babar~\cite{Aubert:2006vb}, Belle~\cite{:2009zv} and CDF~\cite{Aaltonen:2011ja} experiments with much smaller data samples. The CDF experiment has also previously studied the $\phi$ angle. Even with the larger dataset available in this analysis, it is not yet possible to fit the data for all 11 angular terms. Instead, rather than examining the one dimensional projections as has been done in previous analyses,  the angle $\phi$ is transformed such that  

\begin{equation}
\hat{\phi} = 
\begin{cases} \phi + \pi & \text{~if~} \phi < 0
\\
\phi & \text{~otherwise}
\end{cases}
\label{eq:folding}
\end{equation}

\noindent to cancel terms in Eq.~\ref{eq:fullangular} that have either a $\sin\phi$ or a $\cos\phi$ dependence. This provides a simplified angular expression, which contains only $F_{\rm L}$, $A_{\rm FB}$, $S_{3}$ and $A_{9}$,

\begin{equation}
\begin{split}
\frac{1}{\deriv\Gamma/\deriv\qsq} \frac{\deriv^4\Gamma}{\deriv\qsq\,\deriv\cos\theta_{\ell}\,\deriv\cos\theta_{K}\,\deriv\hat{\phi}} = \frac{9}{16\pi} & \left[ \frac{}{} F_{\rm L} \cos^{2}\theta_{K}  + \frac{3}{4} (1-F_{\rm L}) (1-\cos^{2}\theta_{K} ) ~~- \right.  \\
& \left. ~\frac{}{} \,F_{\rm L} \cos^{2}\theta_{K}  (2\cos^{2}\theta_{\ell} -1)  ~~+ \right. \\
& \left. ~\frac{}{} ~\frac{1}{4} (1-F_{\rm L}) (1-\cos^{2}\theta_{K}) (2\cos^{2}\theta_{\ell} -1) ~~+ \right. \\
& \left. ~\frac{}{} ~S_{3} (1-\cos^{2}\theta_{K})(1-\cos^{2}\theta_{\ell}) \cos 2\hat{\phi} ~~+ \right. \\
& \left. ~\frac{}{} ~\frac{4}{3} A_{\rm FB} (1-\cos^{2} \theta_{K}) \cos\theta_{\ell} ~~+  \right. \\
& \left. ~\frac{}{} ~A_{9}(1-\cos^{2}\theta_{K})(1-\cos^{2}\theta_{\ell}) \sin 2\hat{\phi} \frac{}{} ~\right] ~.
\end{split}
\label{eq:fit:noswave}
\end{equation}

\noindent This expression involves the same set of observables that can be extracted from fits to the one-dimensional angular projections.

At large recoil it is also advantageous to reformulate Eq.~\ref{eq:fit:noswave} in terms of the observables $A_{\rm T}^{2}$ and $A_{\rm T}^{\rm Re}$, where  $A_{\rm FB} = \frac{3}{4}\left( 1-F_{\rm L} \right) A_{\rm T}^{\rm Re}$. These so called ``transverse'' observables only depend on a subset of the decay amplitudes (with transverse polarisation of the \Kstarz) and are expected to come with reduced form-factor uncertainties~\cite{Kruger:2005ep, Becirevic:2011bp}. A first measurement of $A_{\rm T}^{2}$ was performed by the CDF experiment~\cite{Aaltonen:2011ja}.

This paper presents a measurement of the differential branching fraction ($\deriv\BF/\deriv\qsq$), $A_{\rm FB}$, $F_{\rm L}$, $S_{3}$ and $A_9$ of the \decay{\Bz}{\Kstarz\mumu} decay in six bins of \qsq. Measurements of the transverse observables $A_{\rm T}^{2}$ and $A_{\rm T}^{\rm Re}$ are also presented. The analysis is based on a dataset, corresponding to 1.0\invfb of integrated luminosity, collected by the \lhcb detector in $\sqrt{s} = 7\tev$ $pp$ collisions in 2011. Section~\ref{sec:detector} describes the experimental setup used in the analyses. Section~\ref{sec:selection} describes the event selection. Section~\ref{sec:selection:backgrounds} discusses potential sources of peaking background. Section~\ref{sec:acceptance} describes the treatment of the detector acceptance in the analysis. Section~\ref{sec:diffbr} discusses the measurement of $\deriv\BF/\deriv\qsq$. The angular analysis of the decay, in terms of $\cos\theta_{\ell}$, $\cos\theta_{K}$ and $\hat{\phi}$, is described in Sec.~\ref{sec:angular}. Finally, a first measurement of the zero-crossing point of $A_{\rm FB}$ is presented in Sec.~\ref{sec:zerocrossing}.

\section{\texorpdfstring{The \lhcb detector}{The LHCb detector}}
\label{sec:detector}

The \lhcb detector~\cite{Alves:2008zz} is a single-arm forward spectrometer, covering the \mbox{pseudorapidity} range $2 < \eta < 5$, that is designed to study $b$ and $c$ hadron decays.  A dipole magnet with a bending power of 4\,Tm and a large area tracking detector provide momentum resolution ranging from 0.4\% for tracks with a momentum of 5\gevc to 0.6\% for a momentum of 100\gevc. A silicon microstrip detector, located around the $pp$ interaction region, provides excellent separation of \B meson decay vertices from the primary $pp$ interaction and impact parameter resolution of 20\mum for tracks with high transverse \mbox{momentum (\pt)}. Two ring-imaging Cherenkov (RICH) detectors~\cite{LHCb-DP-2012-003} provide kaon-pion separation in the momentum range $2-100\gevc$. Muons are identified based on hits created in a system of multiwire proportional chambers interleaved with layers of iron. The LHCb trigger~\cite{LHCb-DP-2012-004} comprises a hardware trigger and a two-stage software trigger that performs a full event reconstruction.

Samples of simulated events are used to estimate the contribution from specific sources of exclusive backgrounds and the efficiency to trigger, reconstruct and select the \decay{\Bz}{\Kstarz\mumu} signal. The simulated $pp$ interactions are generated using
\pythia~6.4~\cite{Sjostrand:2006za} with a specific \lhcb
configuration~\cite{LHCb-PROC-2010-056}.  Decays of hadronic particles
are then described by \evtgen~\cite{Lange:2001uf} in which final state
radiation is generated using \photos~\cite{Golonka:2005pn}. Finally, the \geant toolkit~\cite{Allison:2006ve, *Agostinelli:2002hh} is used to simulate the detector response to the particles produced by  \pythia/\evtgen, as described in
Ref.~\cite{LHCb-PROC-2011-006}.  The simulated samples are corrected for known differences between data and simulation in the \Bz momentum spectrum, the detector impact parameter resolution, particle identification~\cite{LHCb-DP-2012-003} and tracking system performance using control samples from the data.

\section{Selection of signal candidates}
\label{sec:selection}

The \decay{\Bz}{\Kstarz\mumu} candidates are selected from events that have been triggered by a muon with $\pt > 1.5\gevc$, in the hardware trigger. In the first stage of the software trigger, candidates are selected if there is a reconstructed track in the event with high impact parameter ($ > 125\mum$) with respect to one of the primary $pp$ interactions and $\pt > 1.5\gevc$. In the second stage of the software trigger, candidates are triggered on the kinematic properties of the partially or fully reconstructed \Bz candidate~\cite{LHCb-DP-2012-004}.

Signal candidates are then required to pass a set of loose (pre-)selection requirements. Candidates are selected for further analysis if: the \Bz decay vertex is separated from the primary $pp$ interaction; the \Bz candidate impact parameter is small, and the impact parameters of the charged kaon, pion and muons are large, with respect to the primary $pp$ interaction; and the angle between the \Bz momentum vector and the vector between the primary $pp$ interaction and the \Bz decay vertex is small. Candidates are retained if their $\Kp\pim$ invariant mass is in the range $792 < m({\Kp\pim}) < 992\mevcc$. 

A multivariate selection, using a boosted decision tree (BDT)~\cite{Breiman} with the AdaBoost algorithm\cite{AdaBoost}, is applied to further reduce the level of combinatorial background. The BDT is identical to that described in Ref.~\cite{Aaij:2011aa}. It has been trained on a data sample, corresponding to 36\invpb of integrated luminosity, collected by the \lhcb experiment in 2010. A sample of \decay{\Bz}{\Kstarz\jpsi} (\decay{\jpsi}{\mumu}) candidates is used to represent the \decay{\Bz}{\Kstarz\mumu} signal in the BDT training. The decay \decay{\Bz}{\Kstarz\jpsi} is used throughout this analysis as a control channel. Candidates from the \decay{\Bz}{\Kstarz\mumu} upper mass sideband ($5350 < m(\Kp\pim\mumu) < 5600\mevcc$) are used as a background sample. Candidates with invariant masses below the nominal \Bz mass contain a significant contribution from partially reconstructed $B$ decays and are not used in the BDT training or in the subsequent analysis. They are removed by requiring that candidates have $m({\Kp\pim\mumu}) > 5150\mevcc$. The BDT uses predominantly geometric variables, including the variables used in the above pre-selection. It also includes information on the quality of the \Bz vertex and the fit $\chi^{2}$ of the four tracks. Finally the BDT includes information from the RICH and muon systems on the likelihood that the kaon, pion and muons are correctly identified. Care has been taken to ensure that the BDT does not preferentially select regions of \qsq, $\Kp\pim\mumu$ invariant mass or of the $\Kp\pim\mumu$ angular distribution. The multivariate selection retains 78\% of the signal and 12\% of the background that remains after the pre-selection.

\begin{figure}[!tb]
\centering
\includegraphics[scale=0.50]{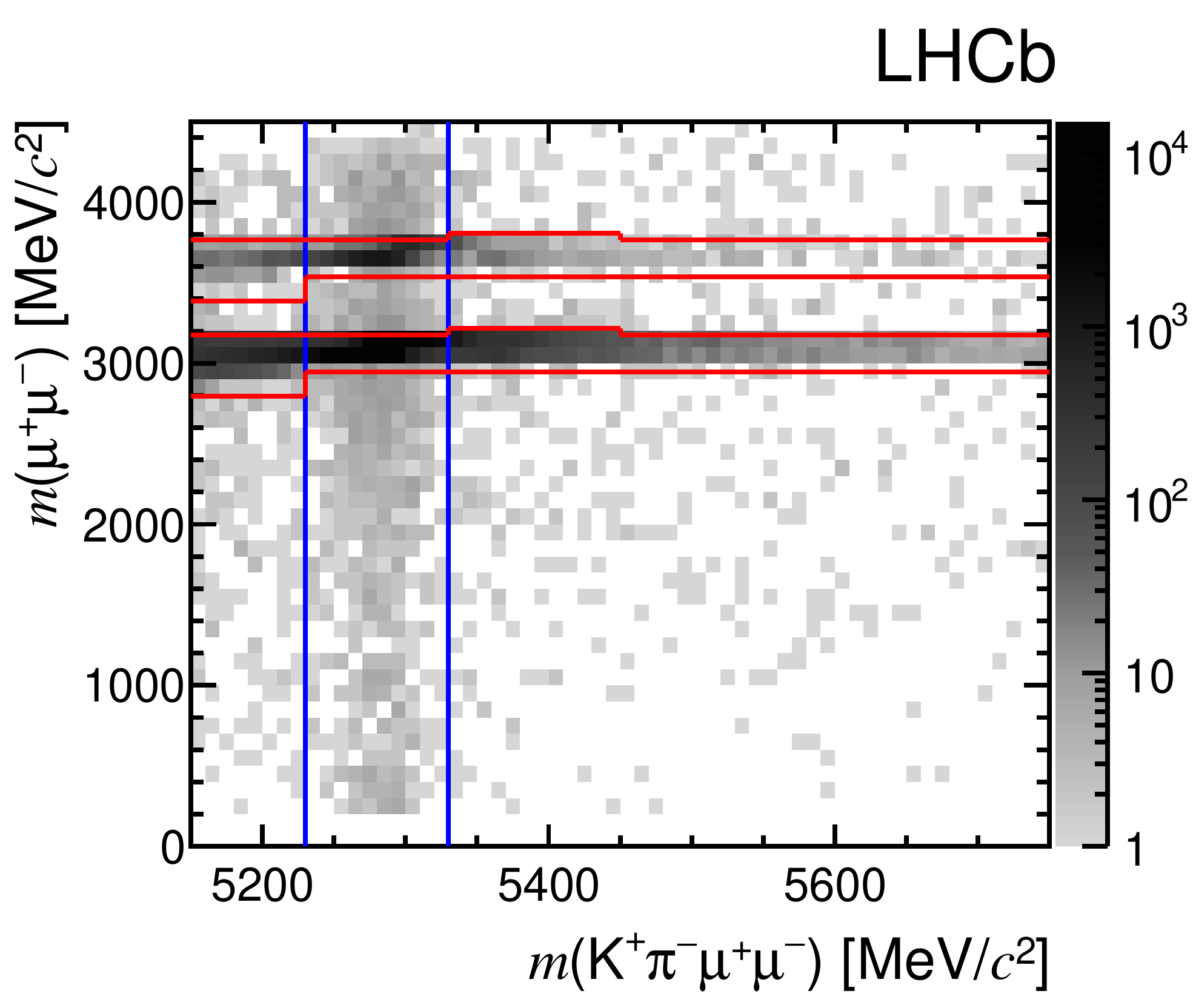}
\caption{{\small Distribution of $\mumu$ versus $\Kp\pim\mumu$ invariant mass of selected \decay{\Bz}{\Kstarz\mumu} candidates. The vertical lines indicate a $\pm 50\mevcc$ signal mass window around the nominal \Bz mass. The horizontal lines indicate the two veto regions that are used to remove \jpsi and \decay{\psitwos}{\mumu} decays. The \decay{\Bz}{\Kstarz\mumu} signal is clearly visible outside of the \jpsi and \decay{\psitwos}{\mumu} windows.} \label{fig:mass:scatter}}
\end{figure}

Figure~\ref{fig:mass:scatter} shows the \mumu versus $\Kp\pim\mumu$ invariant mass of the selected candidates.  The \decay{\Bz}{\Kstarz\mumu} signal, which peaks in $\Kp\pim\mumu$ invariant mass, and populates the full range of the dimuon invariant mass range, is clearly visible.

\section{Exclusive and partially reconstructed backgrounds} 
\label{sec:selection:backgrounds}

Several sources of peaking background have been studied using samples of simulated events, corrected to reflect the difference in particle identification (and misidentification) performance between the data and simulation. Sources of background that are not reduced to a negligible level by the pre- and multivariate-selections are described below.

The decays \decay{\Bz}{\Kstarz\jpsi} and \decay{\Bz}{\Kstarz\psitwos}, where \jpsi and \decay{\psitwos}{\mumu}, are removed by rejecting candidates with $2946 < m({\mumu}) < 3176\mevcc$ and $3586 < m({\mumu}) < 3766\mevcc$. These vetoes are extended downwards by 150\mevcc in $m({\mumu})$ for \decay{\Bz}{\Kstarz\mumu} candidates with masses $5150 < m({\Kp\pim\mumu}) < 5230\mevcc$ to account for the radiative tails of the \jpsi and \psitwos mesons. They are also extended upwards by 25\mevcc for candidates with masses above the \Bz mass to account for the small percentage of \jpsi or \psitwos decays that are misreconstructed at higher masses.  The \jpsi and \psitwos vetoes are shown in Fig.~\ref{fig:mass:scatter}.

The decay  \decay{\Bz}{\Kstarz\jpsi} can also form a source of peaking background if the kaon or pion is misidentified as a muon and swapped with one of the muons from the \jpsi decay. This background is removed by rejecting candidates that have a $\Kp\mun$ or $\pim\mup$ invariant mass (where the kaon or pion is assigned the muon mass) in the range $3036 < m({\mumu}) < 3156\mevcc$ if the kaon or pion can also be matched to hits in the muon stations. A similar veto is applied for the decay \decay{\Bz}{\Kstarz\psitwos}. 

The decay \decay{\Bs}{\phi\mumu}, where \decay{\phi}{\Kp\Km}, is removed by rejecting candidates if the $\Kp\pim$ mass is consistent with originating from a \decay{\phi}{\Kp\Km} decay and the pion is kaon-like according to the RICH detectors. A similar veto is applied to remove \decay{\Lb}{\L^{*}(1520)\mumu} (\decay{\L^{*}(1520)}{p\Km}) decays. 

There is also a source of background from the decay \decay{\Bp}{\Kp\mumu} that appears in the upper mass sideband and has a peaking structure in $\cos\theta_{K}$.  This background arises when a \Kstarz candidate is formed using a pion from the other $B$ decay in the event, and is removed by vetoing events that have a $\Kp\mumu$ invariant mass in the range $5230 < m({\Kp\mumu}) < 5330\mevcc$. The fraction of combinatorial background candidates removed by this veto is small. 

After these selection requirements the dominant sources of peaking background are expected to be from the decays \decay{\Bz}{\Kstarz\jpsi} (where the kaon or pion is misidentified as a muon and a muon as a pion or kaon), \decay{\Bs}{\phi}{\mumu} and \decay{\Bsb}{\Kstarz\mumu} at the levels of $(0.3\pm0.1)\%$, $(1.2\pm 0.5)\%$ and $(1.0\pm1.0)\%$, respectively. The rate of the decay \decay{\Bsb}{\Kstarz\mumu} is estimated using the fragmentation fraction $f_{s}/f_{d}$~\cite{Aaij:2013qqa} and assuming the branching fraction of this decay is suppressed by the ratio of CKM elements $|V_{td}/V_{ts}|^{2}$ with respect to \decay{\Bz}{\Kstarz\mumu}. To estimate the systematic uncertainty arising from the assumed \decay{\Bsb}{\Kstarz\mumu} signal, the expectation is varied by 100\%. Finally, the probability for a decay \decay{\Bz}{\Kstarz\mumu} to be misidentified as \decay{\Bzb}{\Kstarzb\mumu} is estimated to be $(0.85\pm0.02)\%$ using simulated events.

\section{Detector acceptance and selection biases} 
\label{sec:acceptance} 

The geometrical acceptance of the detector, the trigger, the event reconstruction and selection can all bias the angular distribution of the selected candidates. At low \qsq there are large distortions of the angular distribution at extreme values of $\cos\theta_{\ell}$ ($|\cos\theta_{\ell}| \sim 1$). These arise from the requirement that muons have momentum $p \gsim 3\gevc$ to traverse the \lhcb muon system. Distortions are also visible in the $\cos\theta_{K}$ angular distribution. They arise from the momentum needed for a track to reach the tracking system downstream of the dipole magnet, and from the impact parameter requirements  in the pre-selection. The acceptance in $\cos\theta_{K}$ is asymmetric due to the momentum imbalance between the pion and kaon from the \Kstarz decay in the laboratory frame (due to the boost). 

Acceptance effects are accounted for, in a model-independent way by weighting candidates by the inverse of their efficiency determined from simulation. The event weighting takes into account the variation of the acceptance in \qsq to give an unbiased estimate of the observables over the \qsq bin. The candidate weights are normalised such that they have mean 1.0. The resulting distribution of weights in each \qsq bin has a root-mean-square in the range $0.2-0.4$. Less than 2\% of the candidates have weights larger than 2.0. 

The weights are determined using a large sample of simulated three-body \decay{\Bz}{\Kstarz\mumu} phase-space decays. They are determined separately in fine bins of \qsq with widths: $0.1 \gev^2/c^{4}$ for $\qsq < 1\gev^{2}/c^{4}$; $0.2\gev^{2}/c^{4}$ in the range $1 < \qsq < 6\gev^{2}/c^{4}$;  and $0.5\gev^{2}/c^{4}$ for $\qsq > 6\gev^{2}/c^{4}$. The width of the \qsq bins is motivated by the size of the simulated sample and by the rate of variation of the acceptance in \qsq. Inside the \qsq bins, the angular acceptance is assumed to factorise such that $\varepsilon(\cos\theta_{\ell},\cos\theta_{K},\phi) =  \varepsilon(\cos\theta_{\ell})\varepsilon(\cos\theta_{K})\varepsilon(\phi)$. This factorisation is validated at the level of 5\% in the phase-space sample.  The treatment of the event weights is discussed in more detail in Sec.~\ref{sec:angular:statistical}, when determining the statistical uncertainty on the angular observables.

Event weights are also used to account for the fraction of background candidates that were removed in the lower mass ($m(\Kp\pim\mumu) < 5230\mevcc$) and upper mass ($m(\Kp\pim\mumu) > 5330\mevcc$) sidebands by the \jpsi and \psitwos vetoes described in Sec.~\ref{sec:selection:backgrounds} (and shown in Fig.~\ref{fig:mass:scatter}). In each \qsq bin, a linear extrapolation in \qsq is used to estimate this fraction and the resulting event weights.

\section{Differential branching fraction} 
\label{sec:diffbr}

\begin{figure}[!tb]
\centering
\includegraphics[scale=0.36]{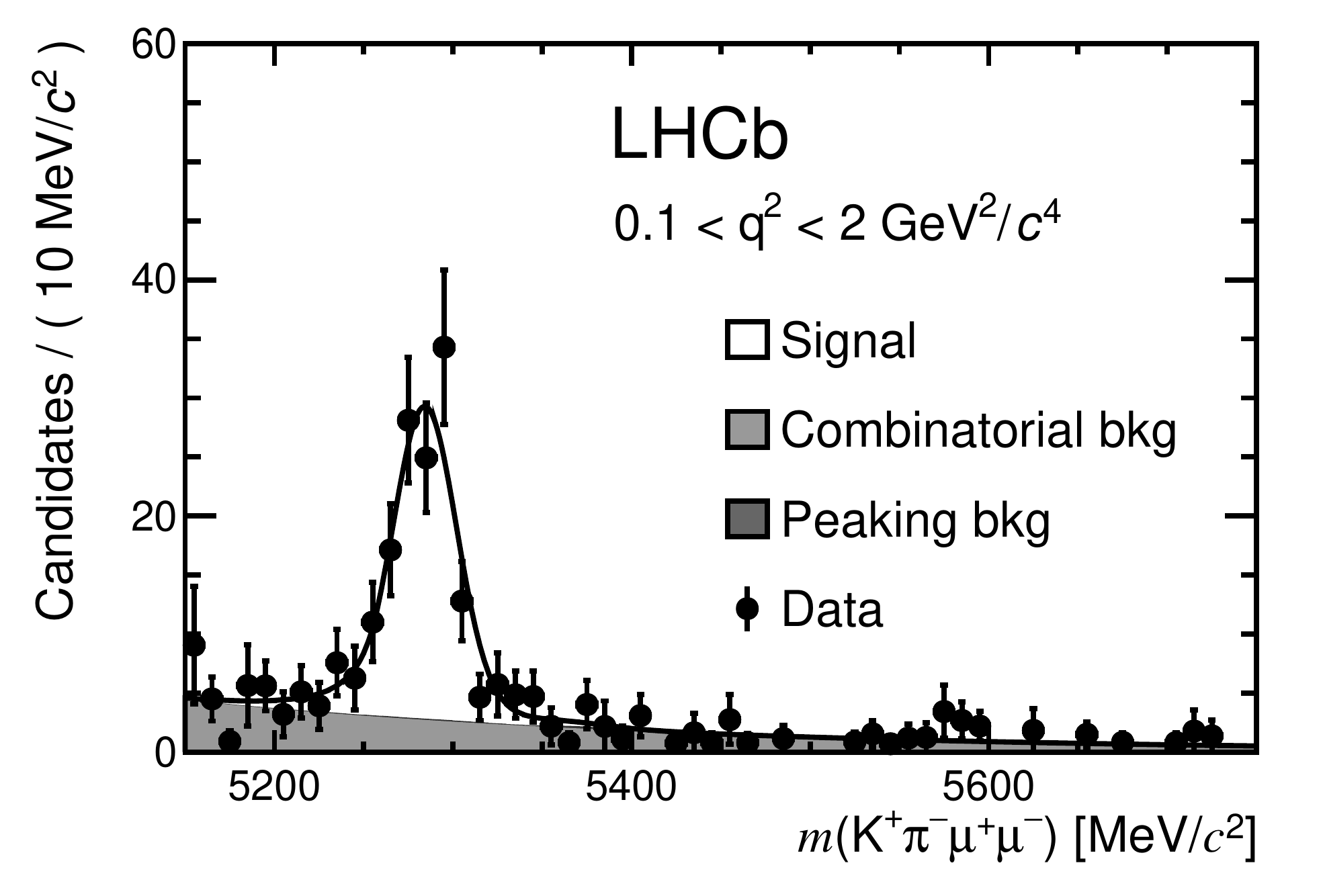}
\includegraphics[scale=0.36]{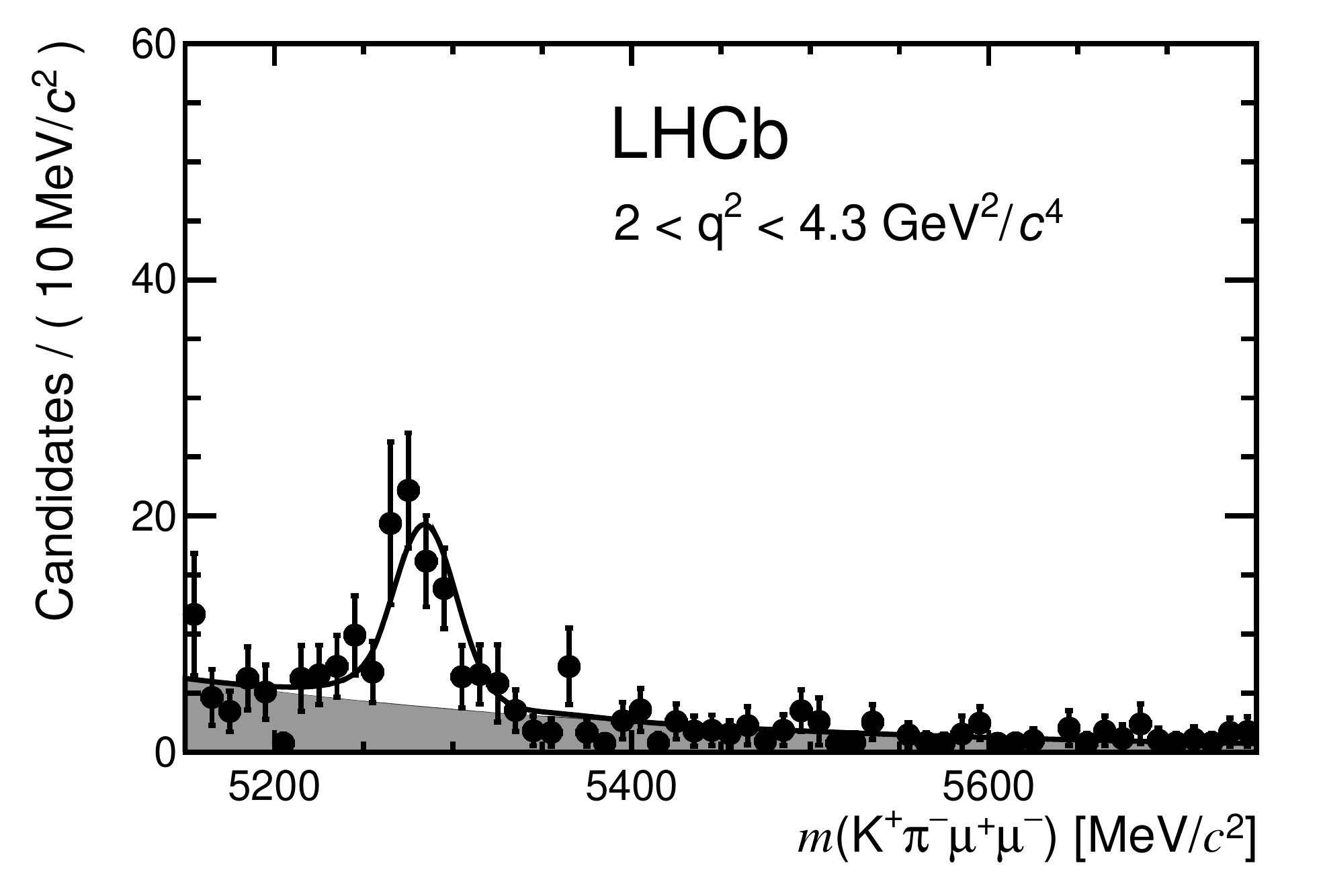} \\
\includegraphics[scale=0.36]{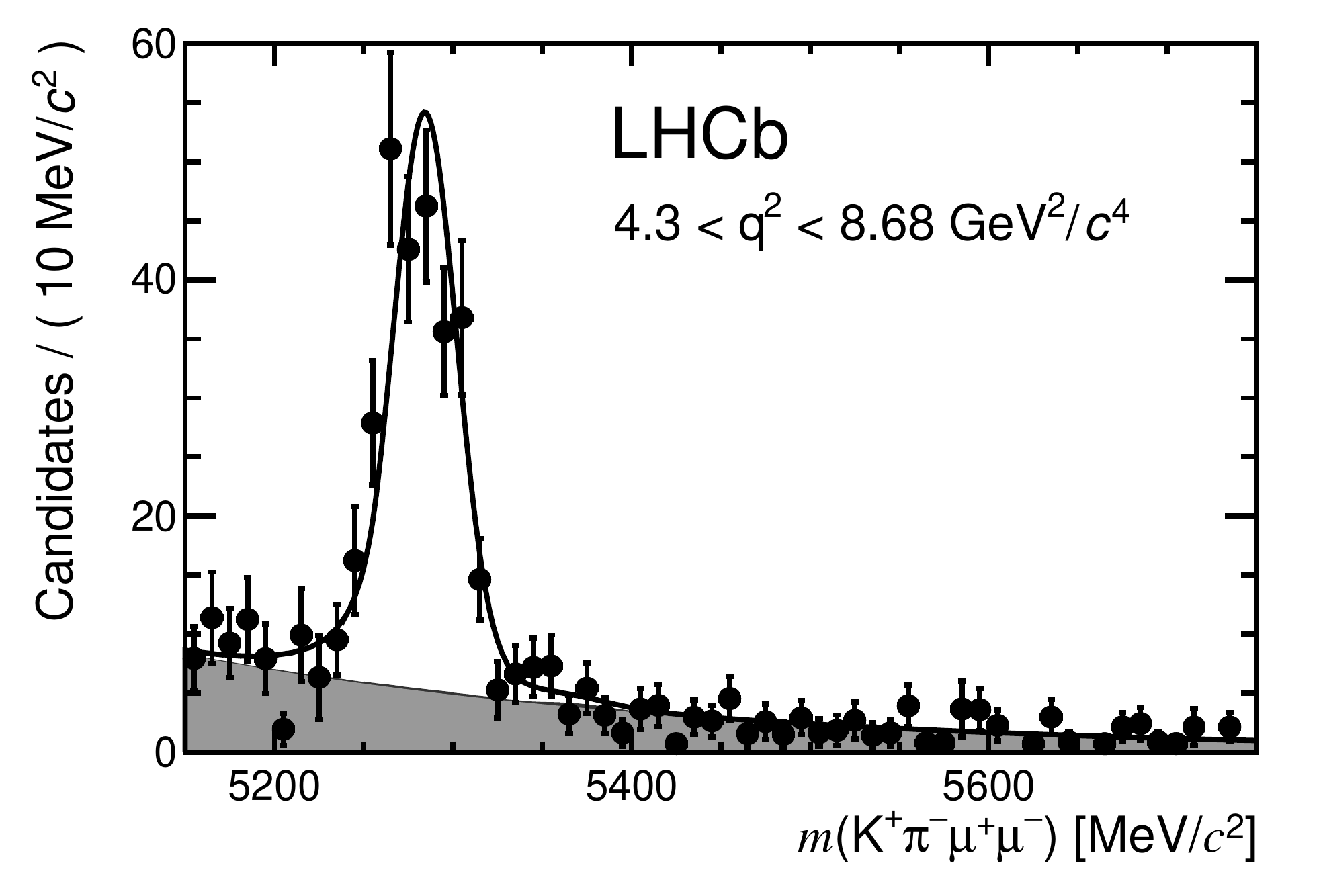}
\includegraphics[scale=0.36]{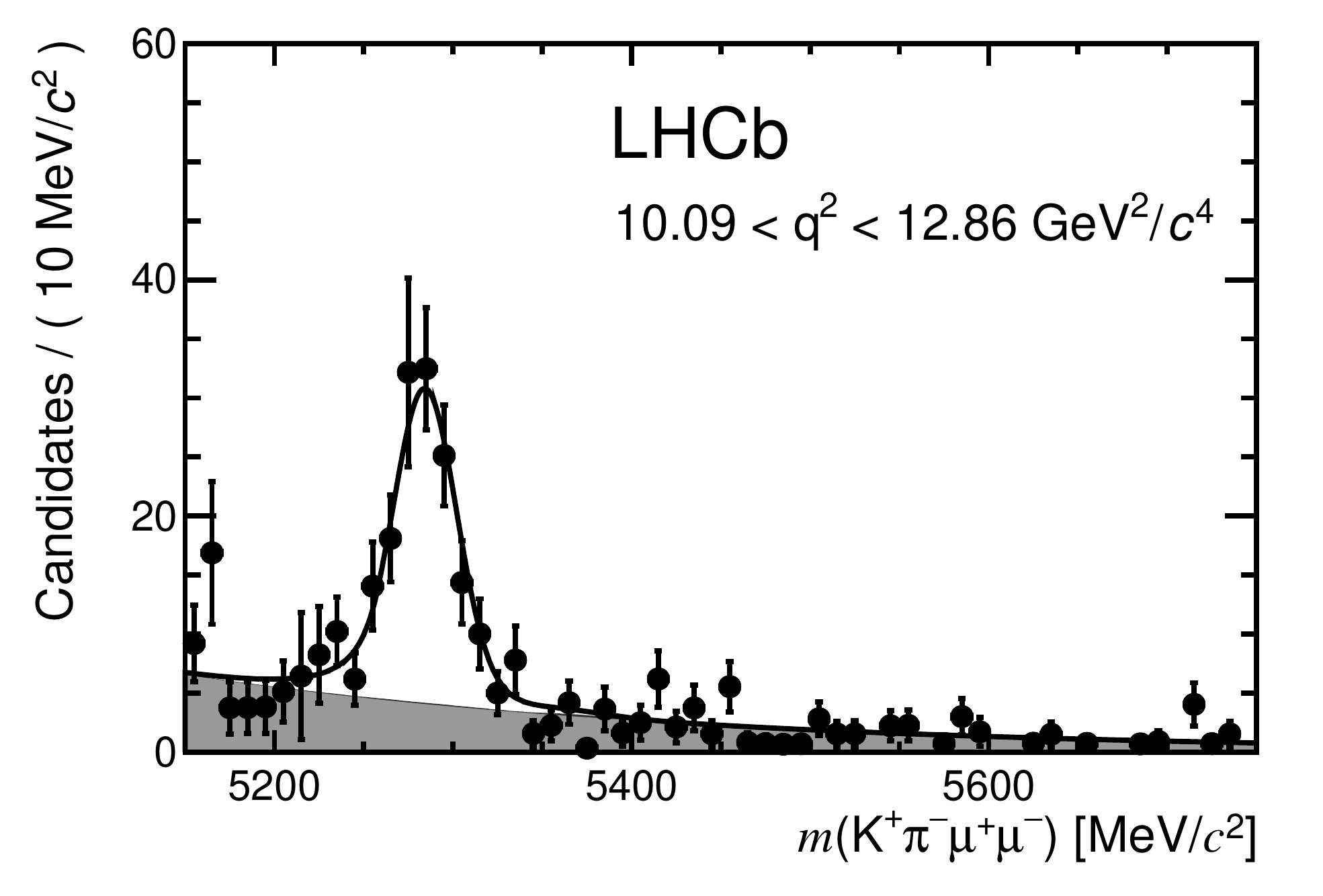} \\
\includegraphics[scale=0.36]{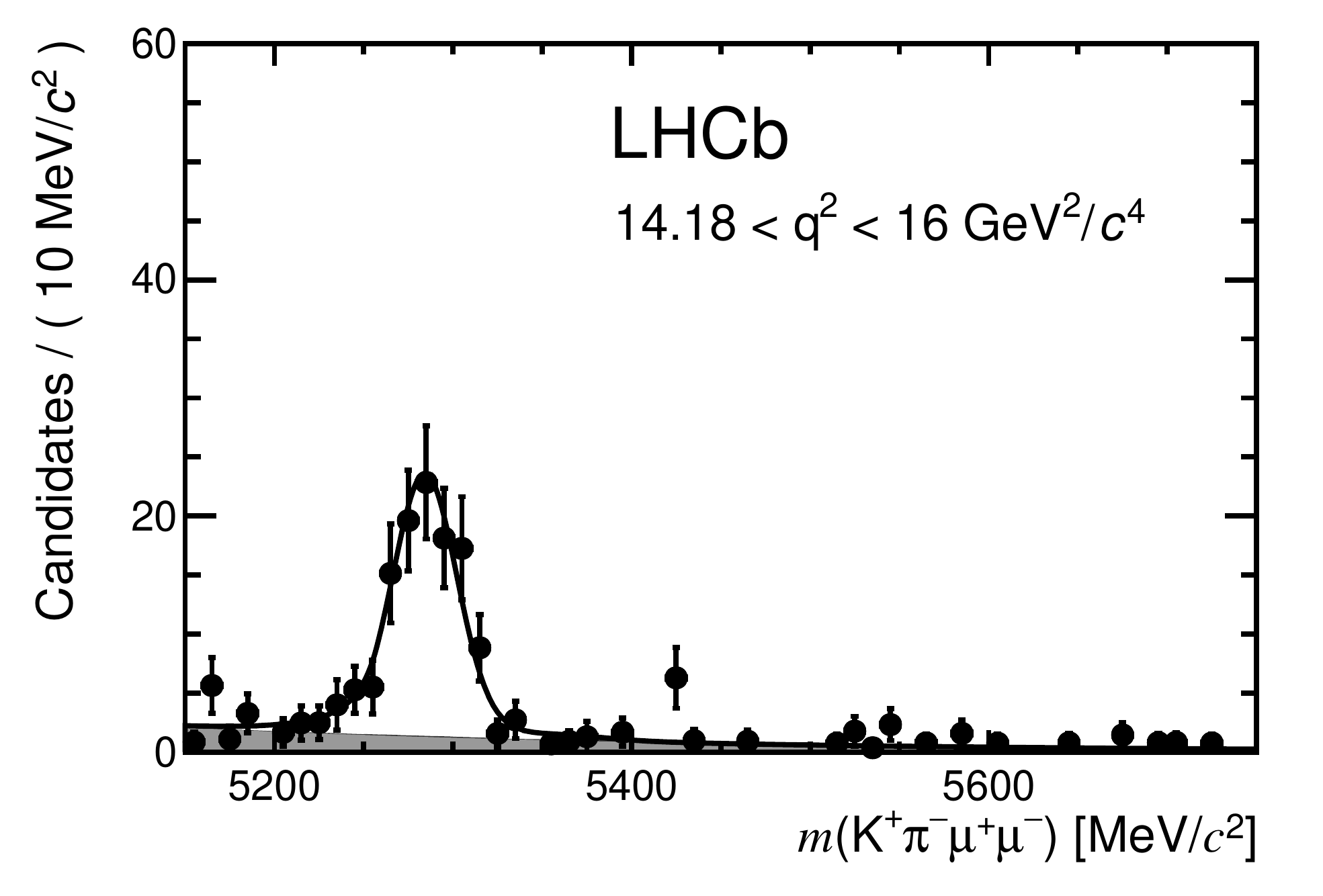}
\includegraphics[scale=0.36]{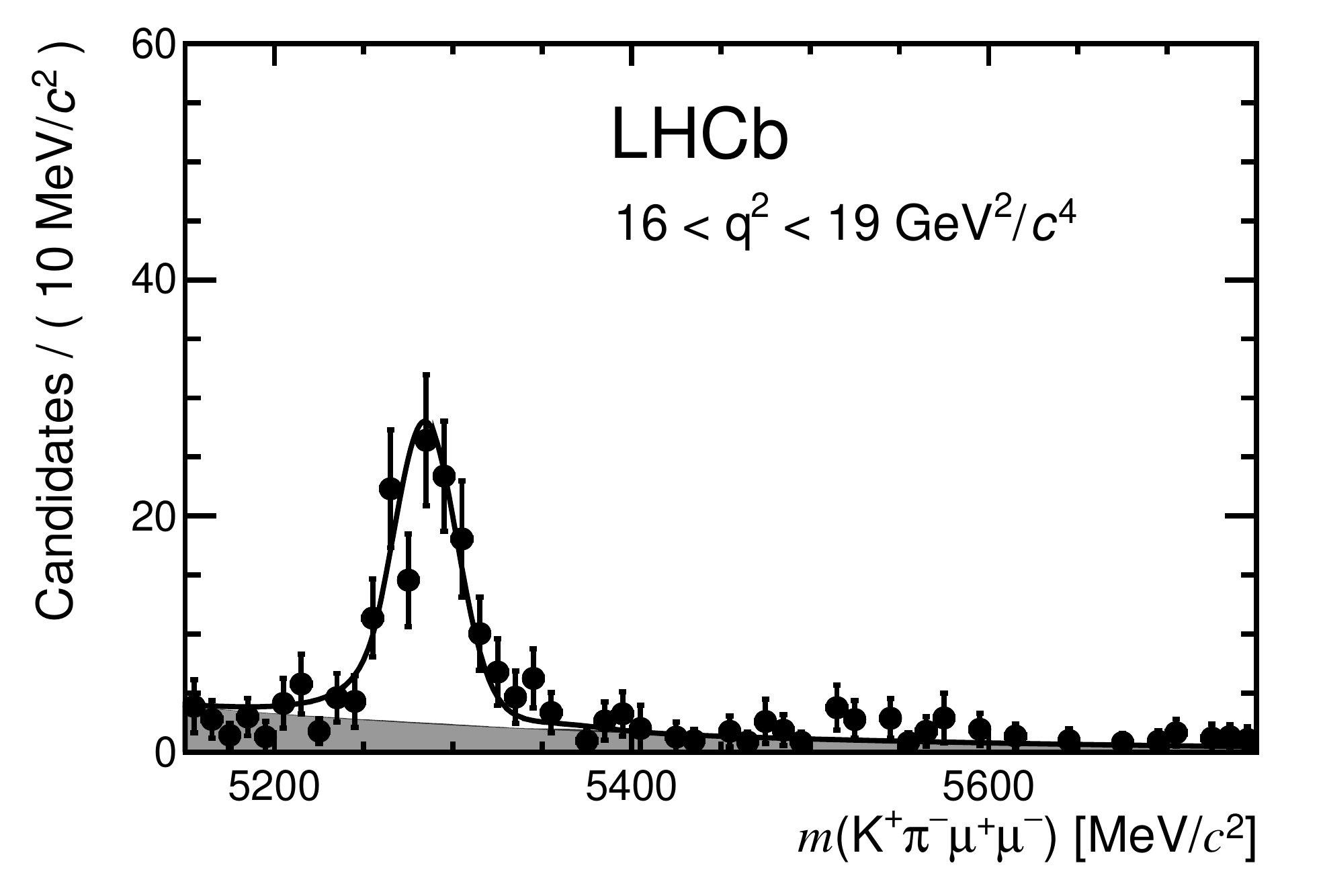} \\
\caption{{\small Invariant mass distributions of $\Kp\pim\mumu$ candidates in the six \qsq bins used in the analysis. The candidates have been weighted to account for the detector acceptance (see text). Contributions from exclusive (peaking) backgrounds are negligible after applying the vetoes described in Sec.~\ref{sec:selection:backgrounds}.} \label{fig:mass:bins}}
\end{figure}

The angular and differential branching fraction analyses are performed in six bins of \qsq, which are the same as those used in Ref.~\cite{:2009zv}. The  $\Kp\pim\mumu$ invariant mass distribution of candidates in these \qsq bins is shown in Fig.~\ref{fig:mass:bins}. 

The number of signal candidates in each of the \qsq bins is estimated by performing an extended unbinned maximum likelihood fit to the $\Kp\pim\mumu$ invariant mass distribution. The signal shape is taken from a fit to the \decay{\Bz}{\Kstarz\jpsi} control sample and is parameterised by the sum of two Crystal Ball~\cite{Skwarnicki:1986xj} functions that differ only by the width of the Gaussian component. The combinatorial background is described by an exponential distribution. The decay \decay{\Bsb}{\Kstarz\mumu}, which forms a peaking background, is assumed to have a shape identical to that of the \decay{\Bz}{\Kstarz\mumu} signal, but shifted in mass by the $\Bs-\Bz$ mass difference~\cite{PDG2012}. Contributions from the decays \decay{\Bs}{\phi\mumu} and \decay{\Bz}{\Kstarz\jpsi} (where the \mun is swapped with the \pim) are also included. The shapes of these backgrounds are taken from samples of simulated events. The sizes of the \decay{\Bsb}{\Kstarz\mumu}, \decay{\Bs}{\phi\mumu} and \decay{\Bz}{\Kstarz\jpsi} backgrounds are fixed with respect to the fitted \decay{\Bz}{\Kstarz\mumu} signal yield according to the ratios described in Sec.~\ref{sec:selection:backgrounds}. These backgrounds are varied to evaluate the corresponding systematic uncertainty. The resulting signal yields are given in Table~\ref{tab:diffbr}. In the full $0.1 < \qsq < 19.0\gev^{2}/c^{4}$ range, the fit yields $883\pm 34$ signal decays.

The differential branching fraction of the decay \decay{\Bz}{\Kstarz\mumu}, in each \qsq bin, is estimated by normalising the \decay{\Bz}{\Kstarz\mumu} yield, $N_{\text{sig}}$, to the total event yield of the \decay{\Bz}{\Kstarz\jpsi} control sample, $N_{\Kstarz\jpsi}$, and correcting for the relative efficiency between the two decays, $\varepsilon_{\Kstarz\jpsi}/\varepsilon_{\Kstarz\mumu}$, 

\begin{equation}
\frac{\deriv \BF}{\deriv \qsq} = \frac{1}{\qsq_{\text{max}}-\qsq_{\text{min}}}\frac{N_{\text{sig}}}{N_{\Kstarz\jpsi}}\frac{\varepsilon_{\Kstarz\jpsi}}{\varepsilon_{\Kstarz\mumu}}\times\BF(\decay{\Bz}{\Kstarz\jpsi})\times\BF(\decay{\jpsi}{\mumu}) ~.
\end{equation}

\noindent The branching fractions $\BF(\decay{\Bz}{\Kstarz\jpsi})$ and $\BF(\decay{\jpsi}{\mumu})$ are $(1.31 \pm 0.03 \pm 0.08) \times 10^{-3}$~\cite{Aubert:2004rz} and $(5.93 \pm 0.06) \times 10^{-2}$~\cite{PDG2012}, respectively.

The efficiency ratio, $\varepsilon_{\Kstarz\jpsi}/\varepsilon_{\Kstarz\mumu}$, depends on the unknown angular distribution of the \decay{\Bz}{\Kstarz\mumu} decay. To avoid making any assumption on the angular distribution, the event-by-event weights described in Sec.~\ref{sec:acceptance} are used to estimate the average efficiency of the \decay{\Bz}{\Kstarz\jpsi} candidates and the signal candidates in each \qsq bin. 

\subsection{Comparison with theory}

\begin{figure}
\centering
\includegraphics[scale=0.55]{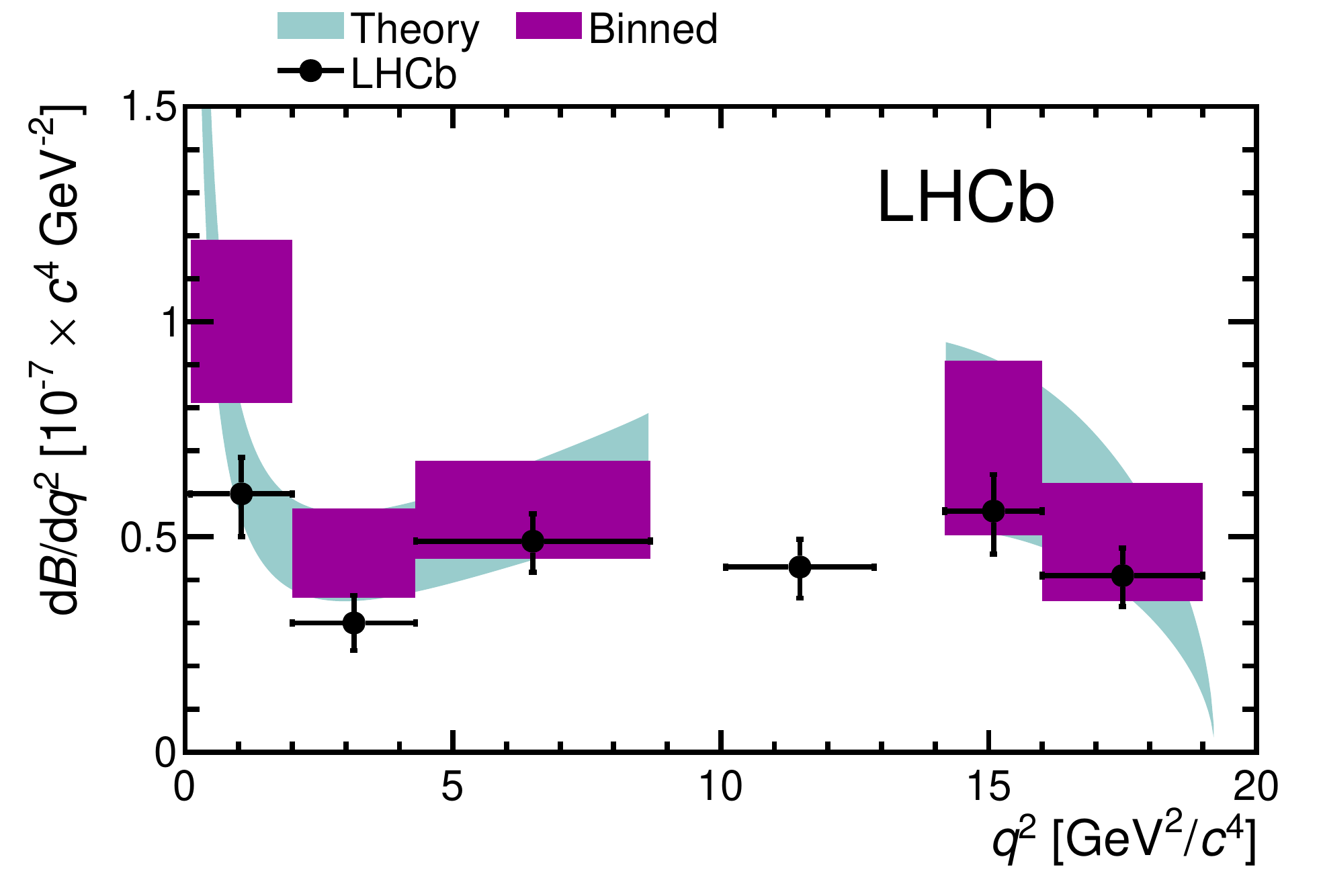}
\caption{{\small Differential branching fraction of the \decay{\Bz}{\Kstarz\mumu} decay as a function of the dimuon invariant mass squared. The data are overlaid with a SM prediction (see text) for the decay (light-blue band). A rate average of the SM prediction across each \qsq bin is indicated by the dark (purple) rectangular regions. No SM prediction is included in the region close to the narrow \ccbar resonances.} \label{fig:diffbr}}
\end{figure}

 The resulting differential branching fraction of the decay \decay{\Bz}{\Kstarz\mumu} is shown in Fig.~\ref{fig:diffbr} and in Table~\ref{tab:diffbr}. The bands shown in Fig.~\ref{fig:diffbr} indicate the theoretical prediction for the differential branching fraction. The calculation of the bands is described in Ref.~\cite{Bobeth:2011gi}.\footnote{A consistent set of SM predictions, averaged over each \qsq bin, have recently also been provided by the authors of Ref.~\cite{Descotes-Genon:2013vna}. }  In the low $q^{2}$ region, the calculations are based on QCD factorisation and soft collinear effective theory (SCET)~\cite{Beneke:2001at}, which profit from  having a heavy \Bz meson and an energetic \Kstarz meson. In the soft-recoil, high \qsq region, an operator product expansion in inverse $b$-quark mass ($1/m_b$) and $1/\sqrt{\qsq}$ is used to estimate the long-distance contributions from quark loops~\cite{Grinstein:2004vb, Beylich:2011aq}.  No theory prediction is included in the region close to the narrow \ccbar resonances (the \jpsi and \psitwos) where the assumptions from QCD factorisation, SCET and the operator product expansion break down. The treatment of this region is discussed in Ref.~\cite{Khodjamirian:2010vf}. The form-factor calculations are taken from Ref.~\cite{PhysRevD.71.014029}. A dimensional estimate is made of the uncertainty on the decay amplitudes from QCD factorisation and SCET of $\mathcal{O}(\Lambda_{\text{QCD}}/m_b)$~\cite{Egede:2008uy}. Contributions from light-quark resonances at large recoil (low \qsq) have been neglected. A discussion of these contributions can be found in Ref.~\cite{Jager:2012uw}. The same techniques are employed in calculations of the angular observables described in Sec.~\ref{sec:angular}.

\begin{table*}
\centering
\caption{{\small Signal yield ($N_{\text{sig}}$) and differential branching fraction ($\deriv\BF/\deriv\qsq$) of the \decay{\Bz}{\Kstarz\mumu} decay in the six \qsq bins used in this analysis. Results are also presented in the $1 < \qsq < 6\gev^{2}/c^{4}$ range where theoretical uncertainties are best controlled. The first and second uncertainties are statistical and systematic. The third uncertainty comes from the uncertainty on the \decay{\Bz}{\Kstarz\jpsi} and \decay{\jpsi}{\mumu} branching fractions. The final uncertainty on $\deriv\BF/\deriv\qsq$ comes from an estimate of the pollution from non-\Kstarz \decay{\Bz}{\Kp\pim\mumu} decays in the $792 < m({\Kp\pim}) < 992\mevcc$ mass window (see Sec.~\ref{sec:systematics:swave}).} \label{tab:diffbr}}
\setlength{\extrarowheight}{2pt}
%\resizebox{\linewidth}{!}{
\begin{tabular}{c|c|c} 
\qsq $(\gev^{2}/c^{4})$ & $N_{\text{sig}}$ & $\deriv\BF/\deriv\qsq$ $(10^{-7} \gev^{-2} c^{4})$ \\
\hline
$\phantom{0}0.10 - \phantom{0}2.00$   & $140 \pm 13$ & $0.60\pm 0.06 \pm 0.05 \pm 0.04 \,^{+0.00}_{-0.05}$ \\
$\phantom{0}2.00 - \phantom{0}4.30$   & $\phantom{0}73 \pm 11$ & $0.30\pm 0.03 \pm 0.03 \pm 0.02 \,^{+0.00}_{-0.02}$ \\
$\phantom{0}4.30 - \phantom{0}8.68$   & $271 \pm 19$ & $0.49\pm 0.04 \pm 0.04 \pm 0.03 \,^{+0.00}_{-0.04}$ \\
$10.09 - 12.86$ & $168 \pm 15$ & $0.43\pm 0.04 \pm 0.04 \pm 0.03 \,^{+0.00}_{-0.03}$ \\
$14.18 - 16.00$ & $115 \pm 12$ & $0.56\pm 0.06 \pm 0.04 \pm 0.04 \,^{+0.00}_{-0.05}$ \\
$16.00 - 19.00$ & $116 \pm 13$ & $0.41\pm 0.04 \pm 0.04 \pm 0.03 \,^{+0.00}_{-0.03}$ \\
\hline
$\phantom{0}1.00 - \phantom{0}6.00$ & $197 \pm 17$ & $0.34\pm 0.03 \pm 0.04 \pm 0.02 \,^{+0.00}_{-0.03}$ \\
\end{tabular}
\end{table*}

\subsection{Systematic uncertainty} 
\label{sec:diffbr:systematics}

The largest sources of systematic uncertainty on the \decay{\Bz}{\Kstarz\mumu} differential branching fraction come from the $\sim 6\%$ uncertainty on the combined \decay{\Bz}{\Kstarz\jpsi} and \decay{\jpsi}{\mumu} branching fractions and from the uncertainty on the pollution of non-\Kstarz decays in the $792 < m({\Kp\pim}) < 992\mevcc$ mass window. The latter pollution arises from decays where the $\Kp\pim$ system is in an S- rather than P-wave configuration. For the decay \decay{\Bz}{\Kstarz\jpsi}, the S-wave pollution is known to be at the level of a few percent~\cite{Aubert:2004cp}. The effect of S-wave pollution on the decay \decay{\Bz}{\Kstarz\mumu} is considered in Sec.~\ref{sec:systematics:swave}. No S-wave correction needs to be applied to the yield of \decay{\Bz}{\Kstarz\jpsi} decays in the present analysis, since the branching fraction used in the normalisation (from Ref.~\cite{Aubert:2004rz}) corresponds to a measurement of the decay \decay{\Bz}{\Kp\pim\jpsi} over the same $m(\Kp\pim)$ window used in this analysis. 

The uncertainty associated with the data-derived corrections to the simulation, which were described in Sec.~\ref{sec:detector}, is estimated to be $1-2\%$. Varying the level of the peaking backgrounds within their uncertainties changes the differential branching fraction by 1\% and this variation is taken as a systematic uncertainty. In the simulation a small variation in the $\Kp\pim\mumu$ invariant mass resolution is seen between \decay{\Bz}{\Kstarz\jpsi} and \decay{\Bz}{\Kstarz\mumu} decays at low and high \qsq, due to differences in the decay kinematics. The maximum size of this variation in the simulation is 5\%. A conservative systematic uncertainty is assigned by varying the mass resolution of the signal decay by this amount in every \qsq bin and taking the deviation from the nominal fit as the uncertainty.

\section{Angular analysis} 
\label{sec:angular} 

This section describes the analysis of the $\cos\theta_{\ell}$, $\cos\theta_{K}$ and $\hat{\phi}$ distribution after applying the transformations that were described earlier. These transformations reduce the full angular distribution from 11 angular terms to one that only depends on four observables: $A_{\rm FB}$, $F_{\rm L}$, $S_{3}$ and $A_{9}$. The resulting angular distribution is given in Eq.~\ref{eq:fit:noswave} in Sec.~\ref{sec:introduction}.

In order for Eq.~\ref{eq:fit:noswave} to remain positive in all regions of the allowed phase space, the observables $A_{\rm FB}$, $F_{\rm L}$, $S_{3}$ and $A_{9}$ must satisfy the constraints

\begin{displaymath}
|A_{\rm FB}| \leq \frac{3}{4}(1-F_{\rm L}) ~,~ |A_{9}| \leq \frac{1}{2}(1-F_{\rm L})~~\text{and}~~ |S_{3}| \leq \frac{1}{2}(1-F_{\rm L}) ~.
\end{displaymath}

\noindent These requirements are automatically taken into account if $A_{\rm FB}$ and $S_3$ are replaced by the theoretically cleaner transverse observables, $A_{\rm T}^{\rm Re}$ and $A_{\rm T}^{2}$,

\begin{displaymath}
A_{\rm FB}  = \frac{3}{4}(1-F_{\rm L})A_{\rm T}^{\rm Re} ~~\text{and}~~ S_{3} = \frac{1}{2}(1-F_{\rm L})A_{\rm T}^{2} ~,
\end{displaymath} 

\noindent which are defined in the range $[-1, 1]$. 

In each of the \qsq bins, $A_{\rm FB}$ ($A_{\rm T}^{\rm Re}$), $F_{\rm L}$, $S_{3}$ ($A_{\rm T}^{2}$) and $A_{9}$ are estimated by performing an unbinned maximum likelihood fit to the $\cos\theta_{\ell}$, $\cos\theta_{K}$ and $\hat{\phi}$ distributions of the \decay{\Bz}{\Kstarz\mumu} candidates. The $\Kp\pim\mumu$ invariant mass of the candidates is also included in the fit to separate between signal- and background-like candidates. The background angular distribution is described using the product of three second-order Chebychev polynomials under the assumption that the background can be factorised into three single angle distributions. This assumption has been validated on the data sidebands ($5350 < m({\Kp\pim\mumu}) < 5600\mevcc$). A dilution factor ($\mathcal{D} = 1-2\omega$) is included in the likelihood fit for $A_{\rm FB}$ and $A_9$, to account at first order for the small probability ($\omega$) for a decay \decay{\Bzb}{\Kstarzb\mumu} to be misidentified as \decay{\Bz}{\Kstarz\mumu}. The value of $\omega$ is fixed to $0.85\%$ in the fit (see Sec.~\ref{sec:selection:backgrounds}).

Two fits to the dataset are performed: one, with the signal angular distribution described by Eq.~\ref{eq:fit:noswave}, to measure $F_{\rm L}$, $A_{\rm FB}$, $S_3$ and $A_9$ and a second replacing $A_{\rm FB}$ and $S_3$ with the observables $A_{\rm T}^{\rm Re}$ and $A_{\rm T}^{2}$. The angular observables vary with \qsq within the \qsq bins used in the analysis. The measured quantities therefore correspond to averages over these \qsq bins. For the transverse observables, where the observable appears alongside $1-F_{\rm L}$ in the angular distribution, the averaging is complicated by the \qsq dependence of both the observable and $F_{\rm L}$. In this case, the measured quantity corresponds to a weighted average of the transverse observable over \qsq, with a weight $(1-F_{\rm L})\deriv\Gamma/\deriv\qsq$.

\subsection{Statistical uncertainty on the angular observables}
\label{sec:angular:statistical}

\begin{figure}
\centering
\includegraphics[scale=0.38]{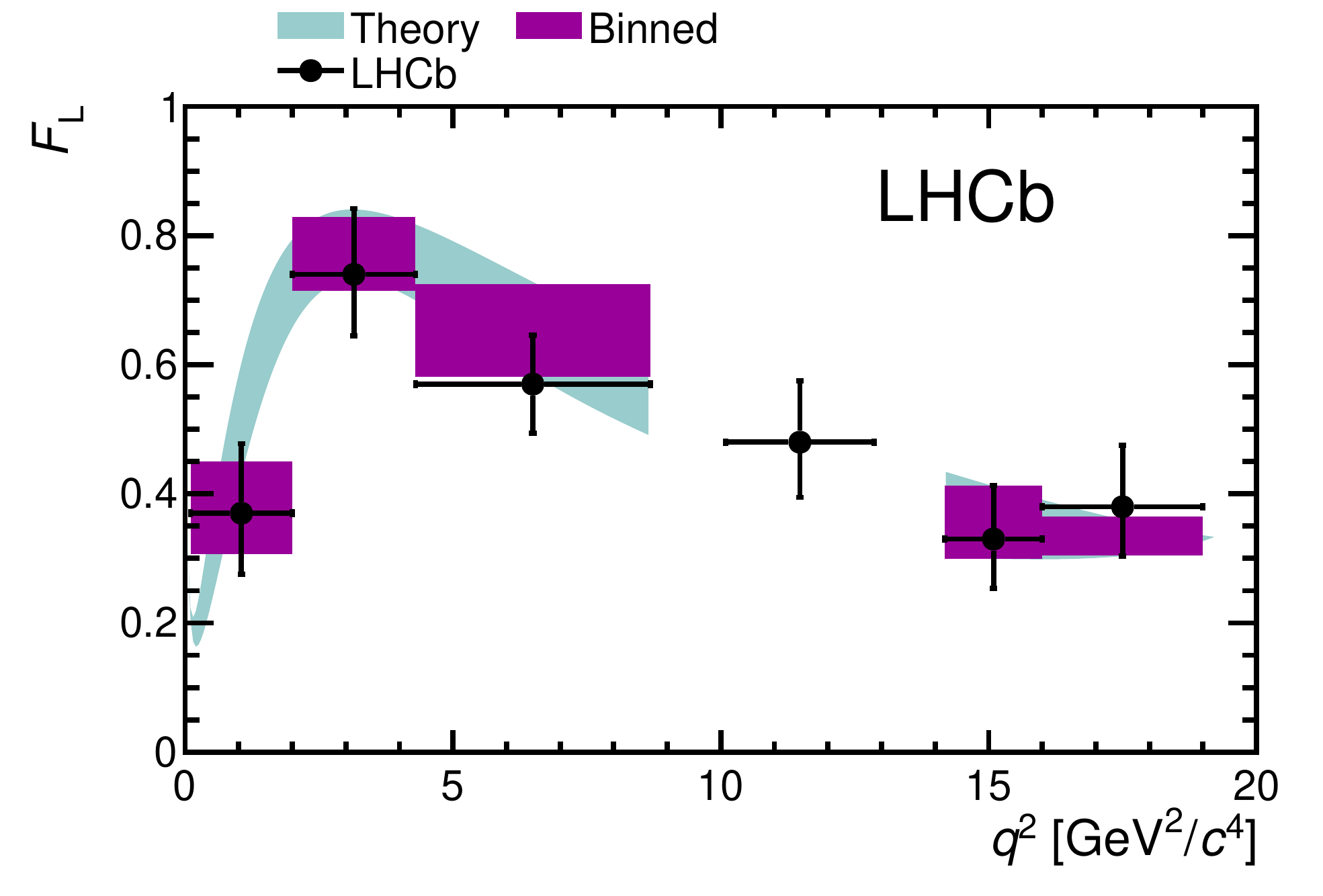} 
\includegraphics[scale=0.38]{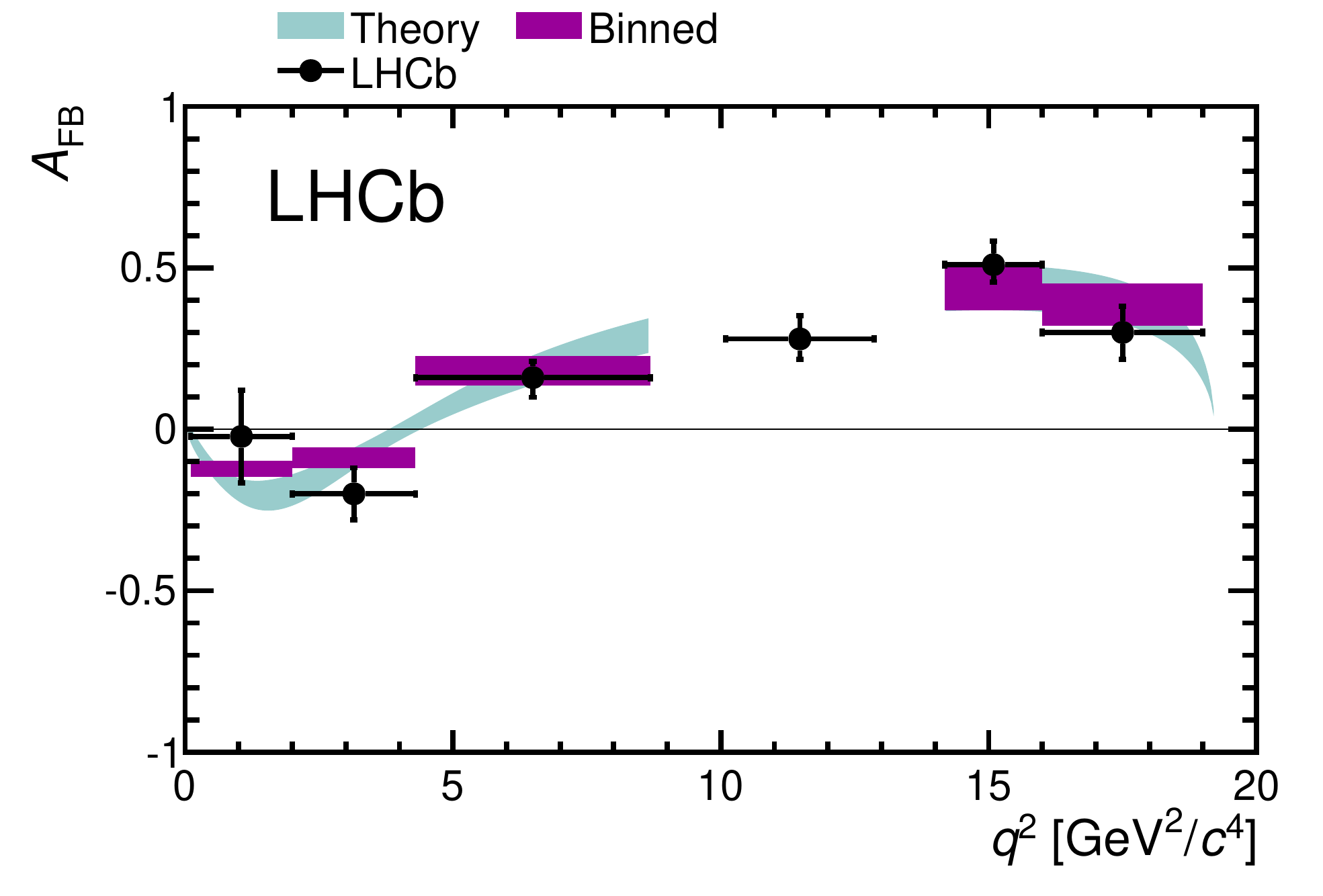} \\
\includegraphics[scale=0.38]{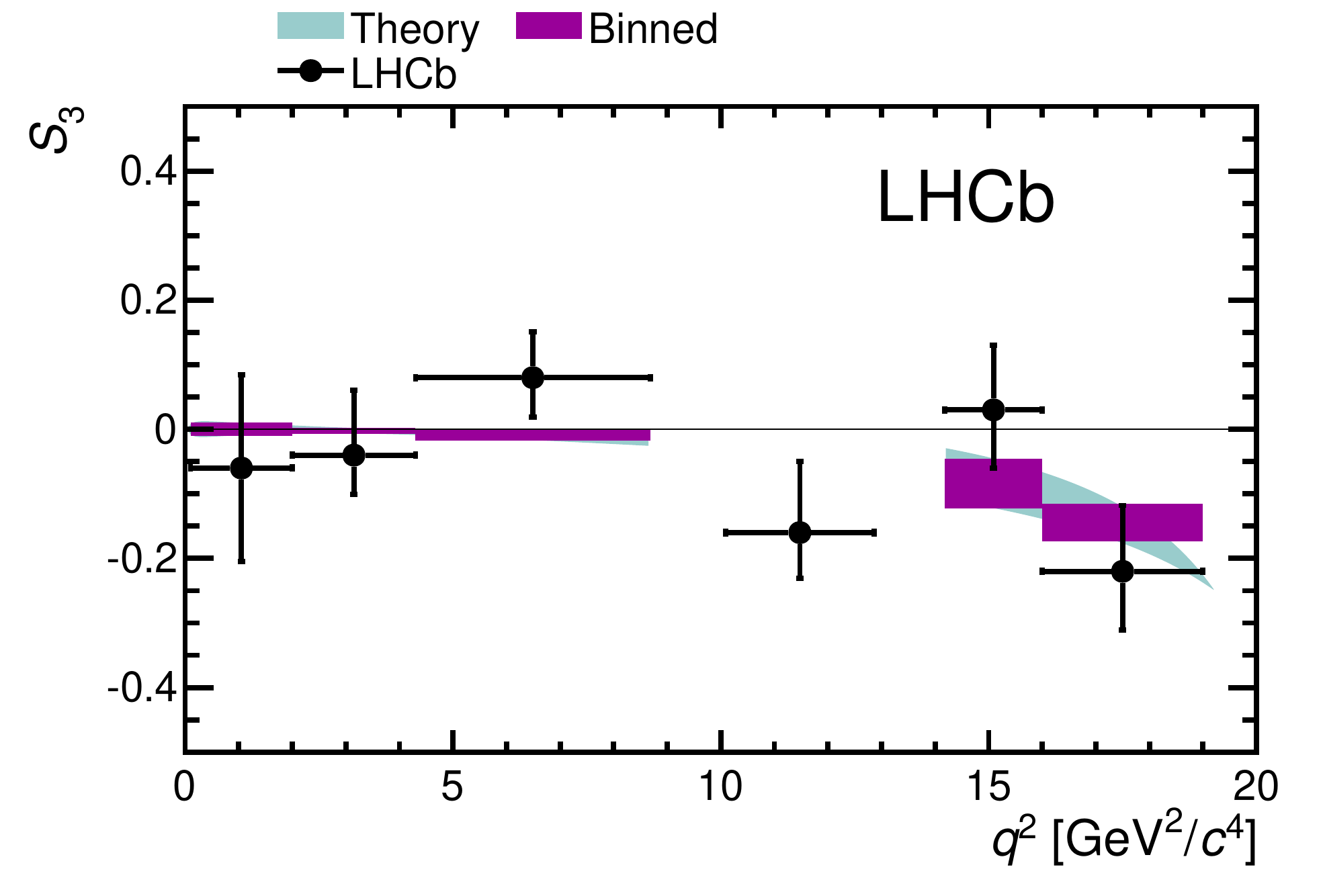} 
\includegraphics[scale=0.38]{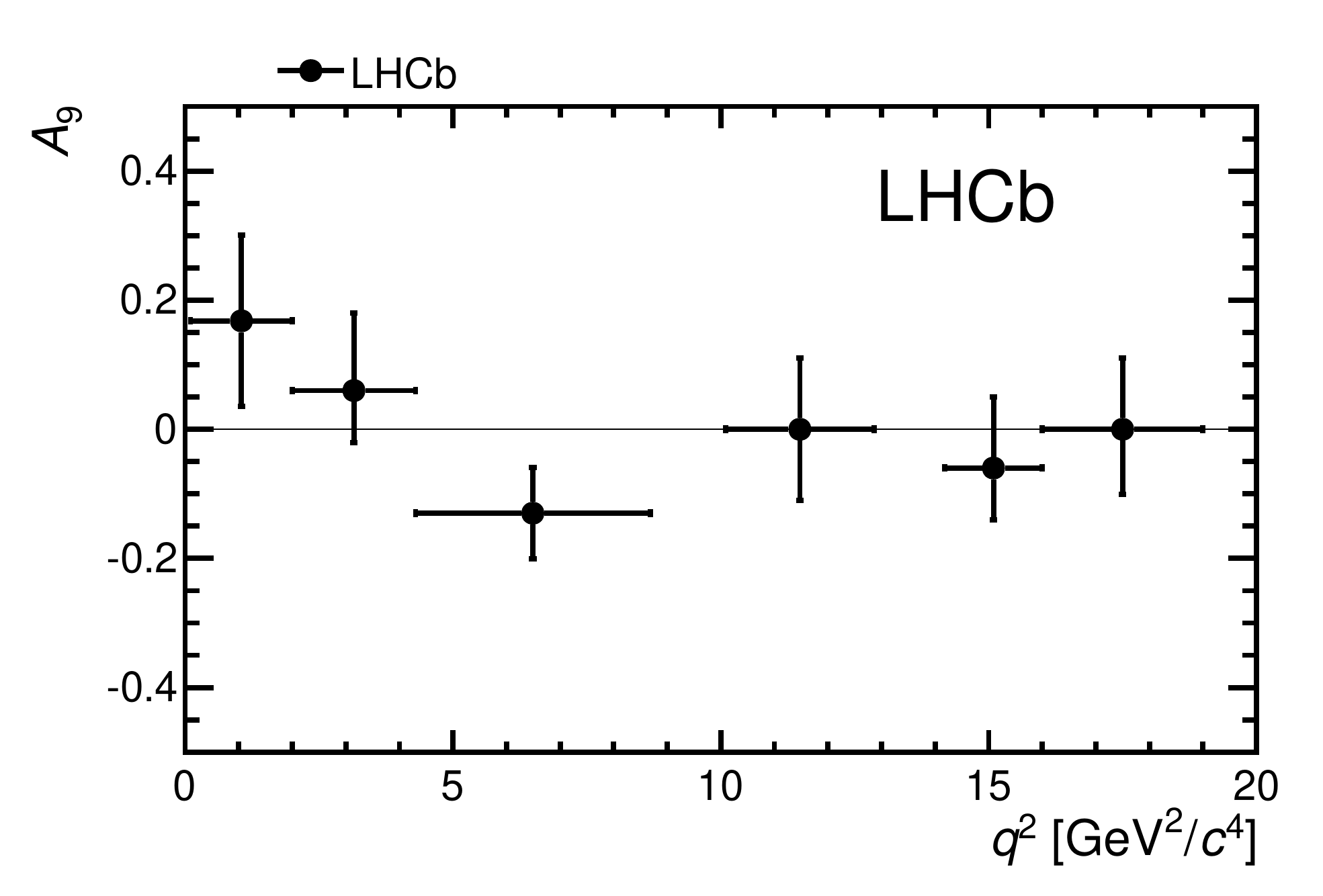}  \\
\caption{{\small Fraction of longitudinal polarisation of the \Kstarz, $F_{\rm L}$, dimuon system forward-backward asymmetry, $A_{\rm FB}$ and the angular observables $S_3$ and $A_9$ from the \decay{\Bz}{\Kstarz\mumu} decay as a function of the dimuon invariant mass squared, \qsq. The lowest \qsq bin has been corrected for the threshold behaviour described in Sec.~\ref{sec:systematics:largerecoil}. The experimental data points overlay the SM prediction described in the text. A rate average of the SM prediction across each \qsq bin is indicated by the dark (purple) rectangular regions. No theory prediction is included for $A_9$, which is vanishingly small in the SM.} \label{fig:standard}}
\end{figure}

\begin{figure}
\centering
\includegraphics[scale=0.38]{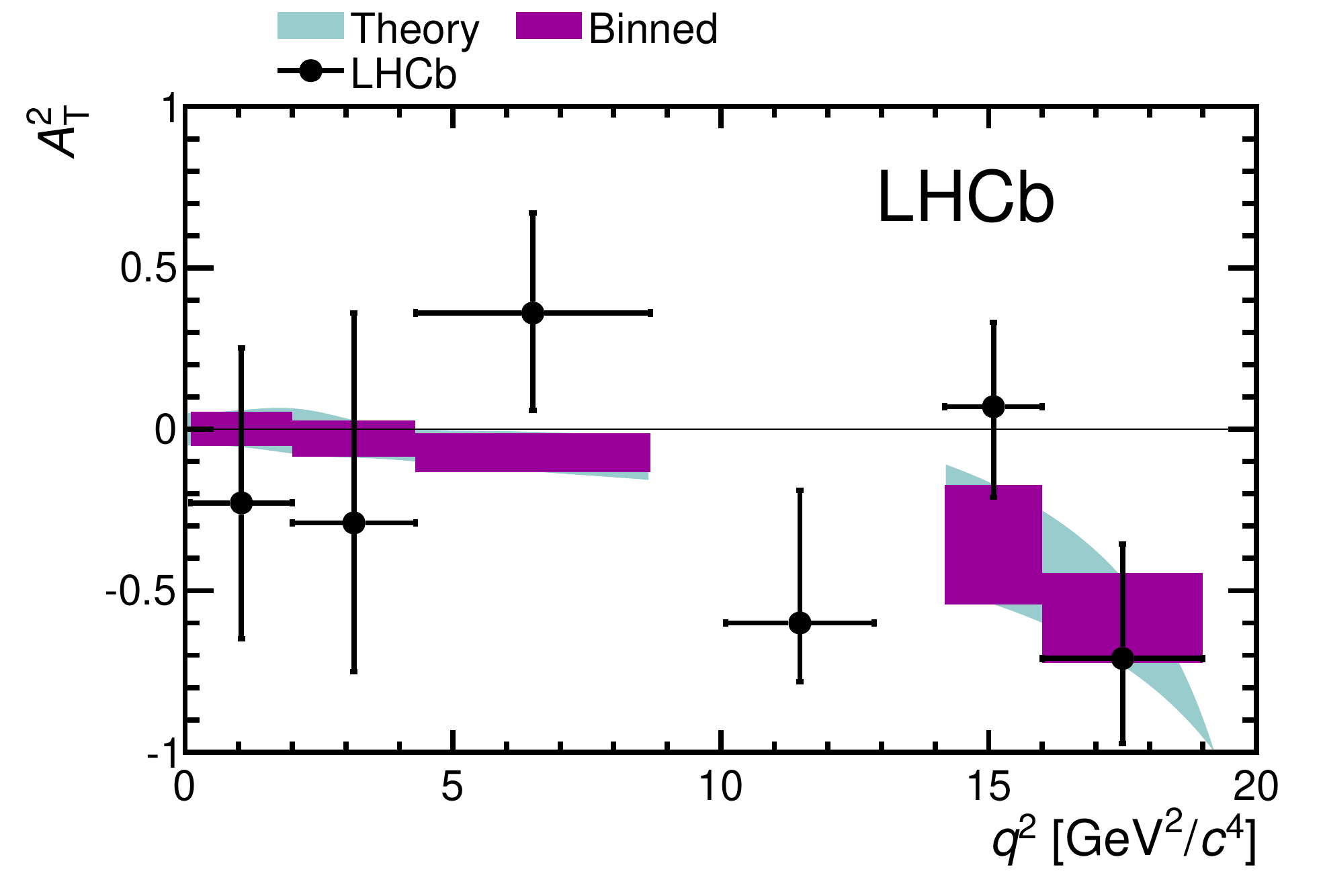} 
\includegraphics[scale=0.38]{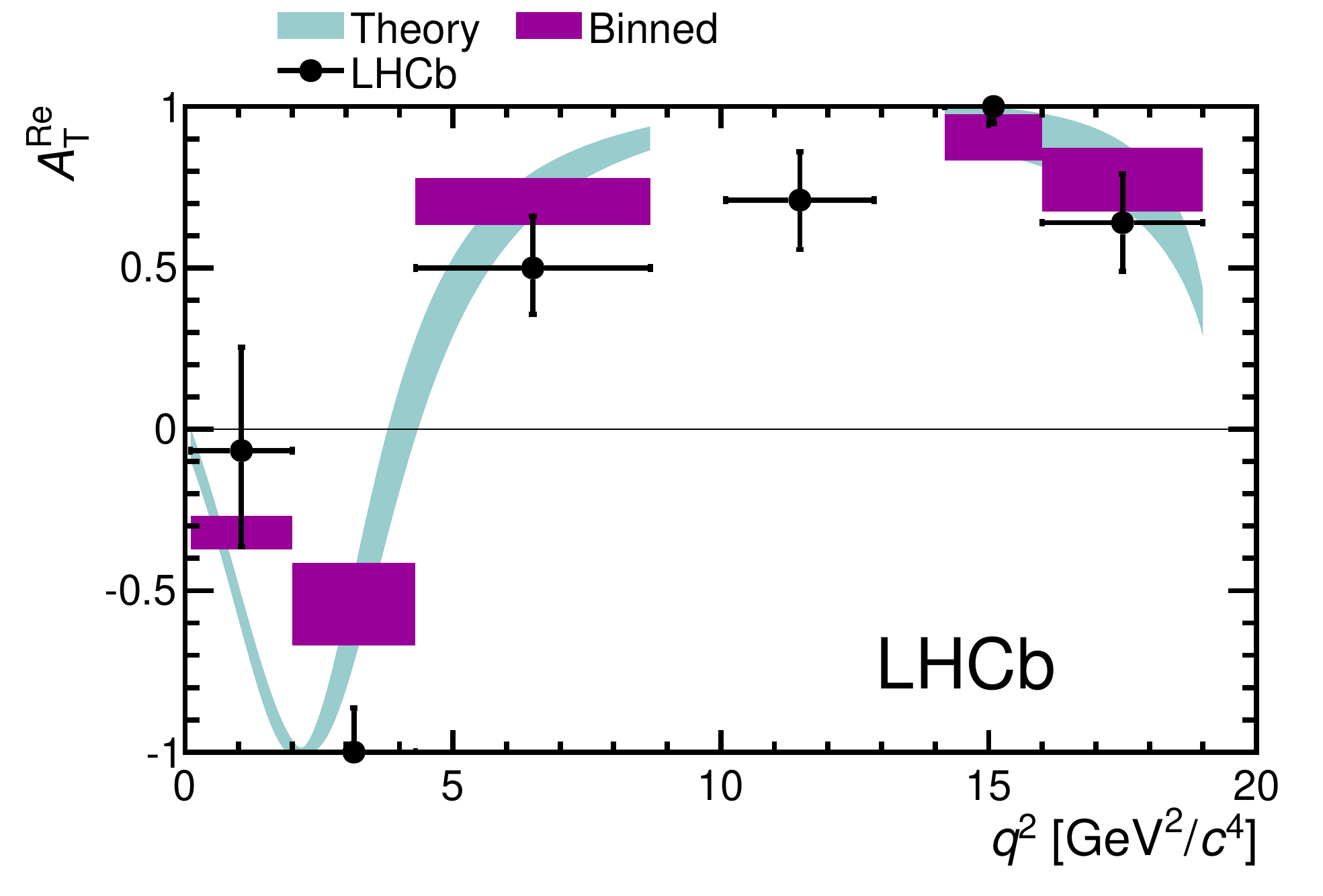} 
\caption{{\small Transverse asymmetries $A_{\rm T}^{2}$ and $A_{\rm T}^{\rm Re}$ as a function of the dimuon invariant mass squared, \qsq, in the  \decay{\Bz}{\Kstarz\mumu} decay. The lowest \qsq bin has been corrected for the threshold behaviour described in Sec.~\ref{sec:systematics:largerecoil}.  The experimental data points overlay the SM prediction that is described in the text. A rate average of the SM prediction across each \qsq bin is indicated by the dark (purple) rectangular regions.} \label{fig:transverse}}
\end{figure}

The results of the angular fits are presented in Table~\ref{tab:angular} and in Figs.~\ref{fig:standard} and \ref{fig:transverse}. The 68\% confidence intervals are estimated using pseudo-experiments and the Feldman-Cousins technique~\cite{Feldman:1997qc}.\footnote{Nuisance parameters are treated according to the ``plug-in'' method (see, for example, Ref.~\cite{woodroofe}).} This avoids any potential bias on the parameter uncertainty that could have otherwise come from using event weights in the likelihood fit or from boundary issues arising in the fitting. The observables are each treated separately in this procedure. For example, when determining the interval on $A_{\rm FB}$, the observables $F_{\rm L}$, $S_3$ and $A_9$ are treated as if they were nuisance parameters. At each value of the angular observable being considered, the maximum likelihood estimate of the nuisance parameters (which also include the background parameters) is used when generating the pseudo-experiments. The resulting confidence intervals do not express correlations between the different observables. The treatment of systematic uncertainties on the angular observables is described in Sec.~\ref{sec:systematics}. 

The final column of Table~\ref{tab:angular} contains the p-value of the SM point in each \qsq bin, which is defined as the probability to observe a difference between the log-likelihood of the SM point compared to the best fit point larger than that seen in the data. They are estimated in a similar way to the Feldman-Cousins intervals by: generating a large ensemble of pseudo-experiments, with all of the angular observables fixed to the central value of the SM prediction; and performing two fits to the pseudo-experiments, one with all of the angular observables fixed to their SM values and one varying them freely. The data are then fitted in a similar manner and the p-value estimated by comparing the ratio of likelihoods obtained for the data to those of the pseudo-experiments. The p-values lie in the range $0.18 - 0.72$ and indicate good agreement with the SM hypothesis. 

As a cross-check, a third fit is also performed in which the sign of the angle $\phi$ for \Bzb decays is flipped to measure $S_9$ in place of $A_9$ in the angular distribution. The term $S_9$ is expected to be suppressed by the size of the strong phases and be close to zero in every \qsq bin. $A_{\rm FB}$ has also been cross-checked by performing a counting experiment in bins of \qsq. A consistent result is obtained in every bin.

\begin{table*}[p]
\centering
\caption{{\small Fraction of longitudinal polarisation of the \Kstarz, $F_{\rm L}$, dimuon system forward-backward asymmetry, $A_{\rm FB}$ and the angular observables $S_3$, $S_9$ and $A_9$ from the \decay{\Bz}{\Kstarz\mumu} decay in the six bins of dimuon invariant mass squared, \qsq, used in the analysis. The lower table includes the transverse observables $A_{\rm T}^{\rm Re}$ and $A_{\rm T}^{2}$, which have reduced form-factor uncertainties. Results are also presented in the $1 < \qsq < 6\gev^{2}/c^{4}$ range where theoretical uncertainties are best controlled. In the large-recoil bin, $0.1 < \qsq < 2.0\gev^{2}/c^{4}$, two results are given to highlight the size of the correction needed to account for changes in the angular distribution that occur when $\qsq \lsim 1\gev^{2}/c^{4}$ (see Sec.~\ref{sec:systematics:largerecoil}). The value of $F_{\rm L}$ is independent of this correction. The final column contains the p-value for the SM point (see text). No SM prediction, and consequently no p-value, is available for the $10.09 < \qsq < 12.86\gev^{2}/c^{4}$ range.} \label{tab:angular}}
\setlength{\extrarowheight}{2pt}
%\resizebox{\linewidth}{!}{
\begin{tabular}{c|c|c|c|c} 
\qsq $(\gev^{2}/c^{4})$ & $F_{\rm L}$ & $A_{\rm FB}$ & $S_{3}$ & $S_{9}$ \\
\hline
$0.10 - 2.00$ & $0.37\,^{+0.10}_{-0.09}\,^{+0.04}_{-0.03}$ & $-0.02\,^{+0.12}_{-0.12}\,^{+0.01}_{-0.01}$ & $-0.04\,^{+0.10}_{-0.10}\,^{+0.01}_{-0.01}$ & $\phantom{-}0.05\,^{+0.10}_{-0.09}\,^{+0.01}_{-0.01}$ \\
(uncorrected) & & & & \\
$0.10 - 2.00$ & $0.37\,^{+0.10}_{-0.09}\,^{+0.04}_{-0.03}$ & $-0.02\,_{-0.13}^{+0.13}\,_{-0.01}^{+0.01}$ &  $-0.05\,_{-0.12}^{+0.12}\,_{-0.01}^{+0.01}$ & $\phantom{-}0.06\,_{-0.12}^{+0.12}\,_{-0.01}^{+0.01}$ \\
(corrected)   & & & & \\
$2.00 - 4.30$ & $0.74\,^{+0.10}_{-0.09}\,^{+0.02}_{-0.03}$ & $-0.20\,^{+0.08}_{-0.08}\,^{+0.01}_{-0.01}$ & $-0.04\,^{+0.10}_{-0.06}\,^{+0.01}_{-0.01}$ & $-0.03\,^{+0.11}_{-0.04}\,^{+0.01}_{-0.01}$ \\
$4.30 - 8.68$ & $0.57\,^{+0.07}_{-0.07}\,^{+0.03}_{-0.03}$ & $\phantom{-}0.16\,^{+0.06}_{-0.05}\,^{+0.01}_{-0.01}$ & $\phantom{-}0.08\,^{+0.07}_{-0.06}\,^{+0.01}_{-0.01}$ & $\phantom{-}0.01\,^{+0.07}_{-0.08}\,^{+0.01}_{-0.01}$ \\
$10.09 - 12.86$ & $0.48\,^{+0.08}_{-0.09}\,^{+0.03}_{-0.03}$ & $\phantom{-}0.28\,^{+0.07}_{-0.06}\,^{+0.02}_{-0.02}$  & $-0.16\,^{+0.11}_{-0.07}\,^{+0.01}_{-0.01}$ & $-0.01\,^{+0.10}_{-0.11}\,^{+0.01}_{-0.01}$\\
$14.18 - 16.00$ & $0.33\,^{+0.08}_{-0.07}\,^{+0.02}_{-0.03}$ & $\phantom{-}0.51\,^{+0.07}_{-0.05}\,^{+0.02}_{-0.02}$ & $\phantom{-}0.03\,^{+0.09}_{-0.10}\,^{+0.01}_{-0.01}$ &  $\phantom{-}0.00\,^{+0.09}_{-0.08}\,^{+0.01}_{-0.01}$\\
$16.00 - 19.00$ & $0.38\,^{+0.09}_{-0.07}\,^{+0.03}_{-0.03}$ & $\phantom{-}0.30\,^{+0.08}_{-0.08}\,^{+0.01}_{-0.02}$ & $-0.22\,^{+0.10}_{-0.09}\,^{+0.02}_{-0.01}$ & $\phantom{-}0.06\,^{+0.11}_{-0.10}\,^{+0.01}_{-0.01}$ \\
\hline
$1.00 - 6.00$ & $0.65\,^{+0.08}_{-0.07}\,^{+0.03}_{-0.03}$ & $-0.17\,^{+0.06}_{-0.06}\,^{+0.01}_{-0.01}$ & $\phantom{-}0.03\,^{+0.07}_{-0.07}\,^{+0.01}_{-0.01}$ & $\phantom{-}0.07\,^{+0.09}_{-0.08}\,^{+0.01}_{-0.01}$ \\

\multicolumn{5}{c}{} \\
\multicolumn{5}{c}{} \\

\qsq $(\gev^{2}/c^{4})$ & $A_{9}$ & $A_{\rm T}^{2}$ & $A_{\rm T}^{\rm Re}$ & p-value \\
\hline
$0.10 - 2.00$ & $\phantom{-}0.12\,_{-0.09}^{+0.09}\,^{+0.01}_{-0.01}$ & $-0.14\,_{-0.30}^{+0.34}\,^{+0.02}_{-0.02}$ & $-0.04\,_{-0.24}^{+0.26}\,^{+0.02}_{-0.01}$ & 0.18 \\
(uncorrected) & & & & \\
$0.10 - 2.00$ & $\phantom{-}0.14\,_{-0.11}^{+0.11}\,_{-0.01}^{+0.01}$ & $-0.19\,_{-0.35}^{+0.40}\,_{-0.02}^{+0.02}$ & $-0.06\,_{-0.27}^{+0.29}\,_{-0.01}^{+0.02}$ & -- \\
(corrected)   & & & & \\
$2.00 - 4.30$ & $\phantom{-}0.06\,_{-0.08}^{+0.12}\,^{+0.01}_{-0.01}$ & $-0.29\,_{-0.46}^{+0.65}\,^{+0.02}_{-0.03}$ & $-1.00\,_{-0.00}^{+0.13}\,^{+0.04}_{-0.00}$ & 0.57 \\
$4.30 - 8.68$ &	$-0.13\,_{-0.07}^{+0.07}\,^{+0.01}_{-0.01}$  & $\phantom{-}0.36\,_{-0.31}^{+0.30}\,^{+0.03}_{-0.03}$ & $\phantom{-}0.50\,_{-0.14}^{+0.16}\,^{+0.01}_{-0.03}$ & 0.71 \\
$10.09 - 12.86$ & $\phantom{-}0.00\,_{-0.11}^{+0.11}\,^{+0.01}_{-0.01}$ &	$-0.60\,_{-0.27}^{+0.42}\,^{+0.05}_{-0.02}$ & $\phantom{-}0.71\,_{-0.15}^{+0.15}\,^{+0.01}_{-0.03}$ &  -- \\
$14.18 - 16.00$ & $-0.06\,_{-0.08}^{+0.11}\,^{+0.01}_{-0.01}$ & $\phantom{-}0.07\,_{-0.28}^{+0.26}\,^{+0.02}_{-0.02}$ & $\phantom{-}1.00\,_{-0.05}^{+0.00}\,^{+0.00}_{-0.02}$ &  0.38 \\ 
$16.00 - 19.00$ & $\phantom{-}0.00\,_{-0.10}^{+0.11}\,^{+0.01}_{-0.01}$ & $-0.71\,_{-0.26}^{+0.35}\,^{+0.06}_{-0.04}$ & $\phantom{-}0.64\,_{-0.15}^{+0.15}\,^{+0.01}_{-0.02}$ & 0.28 \\ 
\hline
$1.00 - 6.00$ & $\phantom{-}0.03\,^{+0.08}_{-0.08}\,^{+0.01}_{-0.01}$ & $\phantom{-}0.15\,_{-0.41}^{+0.39}\,^{+0.03}_{-0.03}$ & $-0.66\,_{-0.22}^{+0.24}\,^{+0.04}_{-0.01}$ & 0.72 \\ 
\end{tabular} 
\end{table*}

\clearpage

\subsection{Angular distribution at large recoil}
\label{sec:systematics:largerecoil}

In the previous section, when fitting the angular distribution, it was assumed that the muon mass was small compared to that of the dimuon system. Whilst this assumption is valid for $\qsq > 2\gev^{2}/c^{4}$, it breaks down in the $0.1 < \qsq < 2.0\gev^{2}/c^{4}$ bin. In this bin, the angular terms receive an additional \qsq dependence, proportional to 

\begin{equation}
 \frac{1 - {4m_{\mu}^{2}}/{\qsq}}{1 + 2 m_{\mu}^{2}/\qsq} ~~\text{or}~~ \frac{(1-4m_{\mu}^{2}/\qsq)^{1/2}}{1 + 2 m_{\mu}^{2}/\qsq} ~,  
\end{equation}

\noindent depending on the angular term $I_j$~\cite{Altmannshofer:2008dz}.

As \qsq tends to zero, these threshold terms become small and reduce the sensitivity to the angular observables. Neglecting these terms leads to a bias in the measurement of the angular observables. Previous analyses by \lhcb, \babar, Belle and CDF have not considered this effect.

The fraction of longitudinal polarisation of the \Kstarz meson, $F_{\rm L}$, is the only observable that is unaffected by the additional terms; sensitivity to $F_{\rm L}$ arises mainly through the shape of the $\cos\theta_{K}$ distribution and this shape remains the same whether the threshold terms are included or not. 

In order to estimate the size of the bias, it is assumed that $A_9$ and $A_{T}^{2}$ are constant over the $0.1 < \qsq < 2\gev^{2}/c^{4}$ region and $A_{\rm T}^{\rm Re}$ rises linearly (with the constraint that $A_{\rm T}^{\rm Re} = 0$ at $\qsq = 0$). Even though $F_{\rm L}$ is in itself unbiased, an assumption needs to be made about the \qsq dependence of $F_{\rm L}$ when determining the bias introduced on the other observables. An empirical model, 

\begin{equation}
\label{eq:flshape}
F_{\rm L}(\qsq)=\frac{a \qsq}{1+a \qsq} ~~,
\end{equation}

\noindent is used. This functional form displays the correct behaviour since it tends to zero as \qsq tends to zero and rises slowly over the \qsq bin, reflecting the dominance of the photon penguin at low \qsq and the transverse polarisation of the photon.

The coefficient $a=0.67\,^{+0.54}_{-0.30}$ is estimated by assigning each (background subtracted) signal candidate a value of $F_{\rm L}$ according to Eq.~\ref{eq:flshape}, averaging $F_{\rm L}$ over the candidates in the \qsq bin and comparing this to the value that is obtained from the fit to the $0.1 < \qsq < 2.0\gev^{2}/c^{4}$ region (in Table~\ref{tab:angular}). Different values of the coefficient $a$ are tried until the two estimates agree. 

To remain model independent, the bias on the angular observables is similarly estimated by summing over the observed candidates. A concrete example of how this is done is given in Appendix~\ref{sec:appendix:largerecoil} for the observable $A_{\rm T}^{2}$.  The typical size of the correction is $10-20\%$. The values of the angular observables, after correcting for the bias, are included in Table~\ref{tab:angular}. A similar factor is also applied to the statistical uncertainty on the fit parameters to scale them accordingly. No systematic uncertainty is assigned to this correction.

The procedure to calculate the size of the bias that is introduced by neglecting the threshold terms has been validated using large samples of simulated events, generated according to the SM prediction and several other scenarios in which large deviations from the SM expectation of the angular observables are possible. In all cases an unbiased estimate of the angular observables is obtained after applying the correction procedure.  Different hypotheses for the \qsq dependence of $F_{\rm L}$, $A_{\rm FB}$ and $A_{\rm T}^{\rm Re}$ do not give large variations in the size of the correction factors.

\subsection{Systematic uncertainties in the angular analysis} 
\label{sec:systematics}

Sources of systematic uncertainty are considered if they introduce either an angular or \qsq dependent bias to the acceptance correction. Moreover, three assumptions have been made that may affect the interpretation of the result of the fit to the $\Kp\pim\mumu$ invariant mass or angular distribution: that $\qsq \gg 4m_{\mu}^{2}$; that there are equal numbers of \Bz and \Bzb decays; and that there is no contribution from non-\Kstarz \decay{\Bz}{\Kp\pim\mumu} decays in the $792 < m({\Kp\pim}) < 992\mevcc$ mass window. The first assumption was addressed in Sec.~\ref{sec:systematics:largerecoil} and no systematic uncertainty is assigned to this correction. The number of \Bz and \Bzb candidates in the data set is very similar~\cite{LHCb:2012kz}. The resulting systematic uncertainty is addressed in Sec.~\ref{sec:systematics:asymmetries}. The final assumption is discussed in Sec.~\ref{sec:systematics:swave} below. 

The full fitting procedure has been tested on \decay{\Bz}{\Kstarz\jpsi} decays. In this larger data sample, $A_{\rm FB}$ is found to be consistent with zero (as expected) and the other observables are in agreement with the results of Ref.~\cite{Aubert:2007hz}. There is however a small discrepancy between the expected parabolic shape of the $\cos\theta_{K}$ distribution and the distribution of the \decay{\Bz}{\Kstarz\jpsi} candidates after weighting the candidates to correct for the detector acceptance. This percent-level discrepancy could point to a bias in the acceptance model. To account for this discrepancy, and any breakdown in the assumption that the efficiencies in $\cos\theta_{\ell}$, $\cos\theta_{K}$ and $\phi$ are independent, systematic variations of the weights are tried in which they are conservatively rescaled by 10\% at the edges of $\cos\theta_{\ell}$, $\cos\theta_{K}$ and $\phi$ with respect to the centre. Several possible variations are explored, including variations that are non-factorisable. The variation which has the largest effect on each of the angular observables is assigned as a systematic uncertainty. The resulting systematic uncertainties are at the level of $0.01 - 0.03$ and are largest for the transverse observables.

The uncertainties on the signal mass model have little effect on the angular observables. Of more importance are potential sources of uncertainty on the background shape. In the angular fit the background is modelled as the product of three second-order polynomials, the parameters of which are allowed to vary freely in the likelihood fit. This model describes the data well in the sidebands. As a cross-check, alternative fits are performed both using higher order polynomials and by fixing the shape of the background to be flat in $\cos\theta_{\ell}$, $\cos\theta_{K}$ and $\hat{\phi}$. The largest shifts in the angular observables occur for the flat background model and are at the level of $0.01 - 0.06$ and $0.02 - 0.25$ for the transverse observables (they are at most 65\% of the statistical uncertainty). These variations are extreme modifications of the background model and are not considered further as sources of systematic uncertainty.

The angular distributions of the decays \decay{\Bs}{\phi\mumu} and \decay{\Bsb}{\Kstarz\mumu} are both poorly known. The decay \decay{\Bsb}{\Kstarz\mumu} is yet to be observed. A first measurement of \decay{\Bs}{\phi\mumu} has been made in Ref.~\cite{LHCb-PAPER-2013-017}. In the likelihood fit to the angular distribution these backgrounds are neglected. A conservative systematic uncertainty on the angular observables is assigned at the level of $\lsim 0.01$ by assuming that the peaking backgrounds have an identical shape to the signal, but have an angular distribution in which each of the observables is either maximal or minimal.

Systematic variations are also considered for the data-derived corrections to the simulated events. For example, the muon identification efficiency, which is derived from data using a tag-and-probe approach with \jpsi decays, is varied within its uncertainty in opposite direction for high ($p > 10\gevc$) and  low ($p < 10 \gevc$) momentum muons. Similar variations are applied to the other data-derived corrections, yielding a combined systematic uncertainty at the level of $0.01 - 0.02$ on the angular observables. The correction needed to account for differences between data and simulation in the \Bz momentum spectrum is small. If this correction is neglected, the angular observables vary by at most 0.01. This variation is associated as a systematic uncertainty.

The systematic uncertainties arising from the variations of the angular acceptance are assessed using pseudo-experiments that are generated with one acceptance model and fitted according to a different model. Consistent results are achieved by varying the event weights applied to the data and repeating the likelihood fit. 

A summary of the different contributions to the total systematic uncertainty can be found in Table~\ref{tab:systematics:summary}. The systematic uncertainty on the angular observables in Table~\ref{tab:angular} is the result of adding these contributions in quadrature. 

\begin{table}
\caption{
{\small Systematic contributions to the angular observables. The values given are the magnitude of the maximum contribution from each source of systematic uncertainty, taken across the six principal \qsq bins used in the analysis.}
\label{tab:systematics:summary}
}
\begin{tabular}{r|rrrrrrr}
Source & $A_{\rm FB}$ & $F_{\rm L}$ & $S_3$ & $S_9$ & $A_9$ & $A_{\rm T}^{2}$ & $A_{\rm T}^{\rm Re}$ \\ 
\hline
Acceptance model    &  $\phantom{<}0.02$ &  $\phantom{<}0.03$ &  $\phantom{<}0.01$ & $<0.01$ & $<0.01$ &  $\phantom{<}0.02$ &  $\phantom{<}0.01$ \\ 
Mass model          & $<0.01$  & $<0.01$ & $<0.01$ & $<0.01$ & $<0.01$ & $<0.01$ & $<0.01$ \\
$\Bz \to \Bzb$ mis-id & $<0.01$  & $<0.01$ & $<0.01$ & $<0.01$ & $\phantom{<}0.01$ & $<0.01$ & $<0.01$ \\
Data-simulation diff. & $\phantom{<}0.01$ &  $\phantom{<}0.03$ &  $\phantom{<}0.01$ & $<0.01$ & $<0.01$ &  $\phantom{<}0.03$ &  $\phantom{<}0.01$ \\
Kinematic reweighting & $<0.01$  & $\phantom{<}0.01$ & $<0.01$ & $<0.01$ & $<0.01$ & $\phantom{<}0.01$ & $<0.01$ \\
Peaking backgrounds & $\phantom{<}0.01$  & $\phantom{<}0.01$ & $\phantom{<}0.01$ & $\phantom{<}0.01$ & $\phantom{<}0.01$ & $\phantom{<}0.01$ & $\phantom{<}0.01$ \\
S-wave              &  $\phantom{<}0.01$ &  $\phantom{<}0.01$ &  $\phantom{<}0.02$ & $\phantom{<}0.01$ & $<0.01$ &  $\phantom{<}0.05$ &  $\phantom{<}0.04$ \\
\Bz-\Bzb asymmetries & $<0.01$  & $<0.01$ & $<0.01$ & $<0.01$ & $<0.01$ & $<0.01$ & $<0.01$ \\
\end{tabular}
\end{table}

\subsubsection{\texorpdfstring{Production, detection and direct ${\boldmath \CP}$ asymmetries}{Production, detection and direct CP asymmetries}} 
\label{sec:systematics:asymmetries}

If the number of \Bz and \Bzb decays are not equal in the likelihood fit then the terms in the angular distribution no longer correspond to pure \CP averages or asymmetries. They instead correspond to admixtures of the two, e.g. 

\begin{equation}
S_{3}^{\text{obs}} \approx S_{3} - A_{3} \left( \mathcal{A}_{\rm CP} + \kappa \mathcal{A}_{\rm P} + \mathcal{A}_{\rm D} \right) ~,
\end{equation} 

\noindent where $\mathcal{A}_{\rm CP}$ is the direct \CP asymmetry between \decay{\Bz}{\Kstarz\mumu} and \decay{\Bzb}{\Kstarzb\mumu} decays;  $\mathcal{A}_{\rm P}$ is the production asymmetry between \Bz and \Bzb mesons, which is diluted by a factor $\kappa$ due to $\Bz-\Bzb$ mixing; and $\mathcal{A}_{\rm D}$ is the detection asymmetry between the \Bz and \Bzb decays (which might be non-zero due to differences in the interaction cross-section with matter between \Kp and \Km mesons). In practice, the production and detection asymmetries are small in \lhcb and $\mathcal{A}_{\rm CP}$ is measured to be $\mathcal{A}_{\rm CP} = -0.072 \pm 0.040 \pm 0.005$~\cite{LHCb:2012kz}, which is consistent with zero. Combined with the expected small size of the \CP asymmetry or \CP-averaged counterparts of the angular observables measured in this analysis, this reduces any systematic bias to $< 0.01$.

\subsubsection{Influence of S-wave interference on the angular distribution}
\label{sec:systematics:swave}

The presence of a non-\Kstarz \decay{\Bz}{\Kp\pim\mumu} component, where the $\Kp\pim$ system is in an S-wave configuration, modifies Eq.~\ref{eq:fit:noswave} to 

\begin{equation}
\begin{split}
\frac{1}{\deriv\Gamma'/\deriv\qsq} \frac{\deriv^4\Gamma'}{\deriv\qsq\,\deriv\cos\theta_{\ell}\,\deriv\cos\theta_{K}\,\deriv\hat{\phi}} = & ~(1-F_{\rm S}) \left[ \frac{1}{\deriv\Gamma/\deriv\qsq} \frac{\deriv^4\Gamma}{\deriv\qsq\,\deriv\cos\theta_{\ell}\,\deriv\cos\theta_{K}\,\deriv\hat{\phi}} \right]  \\ 
&  ~~\, + \frac{9}{16\pi}\left[ \frac{2}{3} F_{\rm S} ( 1 - \cos^{2}\theta_{\ell} ) + \frac{4}{3} A_{\rm S} \cos\theta_{K} ( 1 - \cos^{2}\theta_{\ell} ) \right]~, \\
\end{split}
\label{eq:fit:added}
\end{equation}

\noindent where $F_{\rm S}$ is the fraction of \decay{\Bz}{\Kp\pim\mumu} S-wave in the $792 < m({\Kp\pim}) < 992\mevcc$ window. The partial width, $\Gamma'$, is the sum of the partial widths for the \decay{\Bz}{\Kstarz\mumu} decay and the \decay{\Bz}{\Kp\pim\mumu} S-wave. A forward-backward asymmetry in $\cos\theta_{K}$, $A_{\rm S}$, arises due to the interference between the longitudinal amplitude of the \Kstarz and the S-wave amplitude~\cite{Lu:2011jm,Becirevic:2012dp,Blake:2012mb,Matias:2012qz}.

The S-wave is neglected in the results given in Table~\ref{tab:angular}. To estimate the size of the S-wave component, and the impact it might have on the \decay{\Bz}{\Kstarz\mumu} angular analysis, the phase shift of the  \Kstarz Breit-Wigner function around the \Kstarz pole mass is exploited. Instead of measuring $F_{\rm S}$ directly, the average value of $A_{\rm S}$ is measured in two bins of $\Kp\pim$ invariant mass,  one below and one above the  \Kstarz  pole mass. If the magnitude and phase of the S-wave amplitude are assumed to be independent of the $\Kp\pim$ invariant mass in the range $792 < m({\Kp\pim}) < 992\mevcc$, and the P-wave amplitude is modelled by a Breit-Wigner function, the two $A_{\rm S}$ values can then be used to determine the real and imaginary components of the S-wave amplitude (and $F_{\rm S}$).\footnote{In the decay \decay{\Bz}{\Kstarz\mumu} there are actually two pairs of amplitudes involved, left- and right-handed longitudinal amplitudes and left- and right-handed S-wave amplitudes (where the handedness refers to the chirality of the dimuon system). In order to exploit the interference and determine $F_{\rm S}$ it is assumed that the phase difference between the two left-handed amplitudes is the same as the difference between the two right-handed amplitudes, as expected from the expression for the amplitudes in Refs.~\cite{Lu:2011jm,Becirevic:2012dp}.}

For a small S-wave amplitude, the pure S-wave contribution, $F_{\rm S}$, to Eq.~\ref{eq:fit:added} has only a small effect on the angular distribution. The magnitude of $A_{\rm S}$ arising from the interference between the S- and P-wave can however still be sizable and this information is exploited by this phase-shift method.  The method, described above, is statistically more precise than fitting Eq.~\ref{eq:fit:added} directly for $A_{\rm S}$ and $F_{\rm S}$ as uncorrelated variables. For the \decay{\Bz}{\Kstarz\jpsi} control mode, the gain in statistical precision is approximately a factor of three. 

Due to the limited  number of signal candidates that are available in each of the \qsq bins, the bins are merged in order to estimate the S-wave fraction. In the range $0.1 < \qsq < 19\gev^{2}/c^{4}$, $F_{\rm S} = 0.03 \pm 0.03$, which corresponds to an upper limit of $F_{\rm S}<0.04$ at $68\%$ confidence level (CL). The procedure has also been performed in  the region $1 < \qsq < 6\gev^{2}/c^{4}$, where both $F_{\rm L}$ and $F_{\rm S}$ are expected to be enhanced. This gives  $F_{\rm S} =0.04 \pm 0.04$ and an upper limit of $F_{\rm S} < 0.07$ at $68\%$ CL.  In order to be conservative, $F_{\rm S} = 0.07$ is used to estimate a systematic uncertainty on the differential branching fraction and angular analyses. The \decay{\Bz}{\Kstarz\jpsi} data has been used to validate the method.

For the differential branching fraction analysis, $F_{\rm S}$ scales the observed branching fraction by up to 7\%. For the angular analysis, $F_{\rm S}$ dilutes $A_{\rm FB}$, $S_{3}$ and $A_{9}$. The impact on $F_{\rm L}$ however, is less easy to disentangle. To assess the possible size of a systematic bias, pseudo-experiments have been carried out generating with, and fitting without, the S-wave contribution in the likelihood fit. The typical bias on the angular observables due to the S-wave is $0.01-0.03$.

\section{Forward-backward asymmetry zero-crossing point} 
\label{sec:zerocrossing} 

In the SM, $A_{\rm FB}$ changes sign at a well defined value of \qsq, $q^{2}_{0}$, whose prediction is largely free from form-factor uncertainties~\cite{Ali:1999mm}. It is non-trivial to estimate $q^2_0$ from the angular fits to the data in the different \qsq bins, due to the large size of the bins involved. Instead, $A_{\rm FB}$ can be estimated by counting the number of forward-going ($\cos\theta_{\ell} > 0$) and backward-going ($\cos\theta_{\ell} < 0$) candidates and $q^{2}_{0}$ determined from the resulting distribution of $A_{\rm FB}(\qsq)$.

\begin{figure}[!htb]
\centering
\includegraphics[scale=0.38]{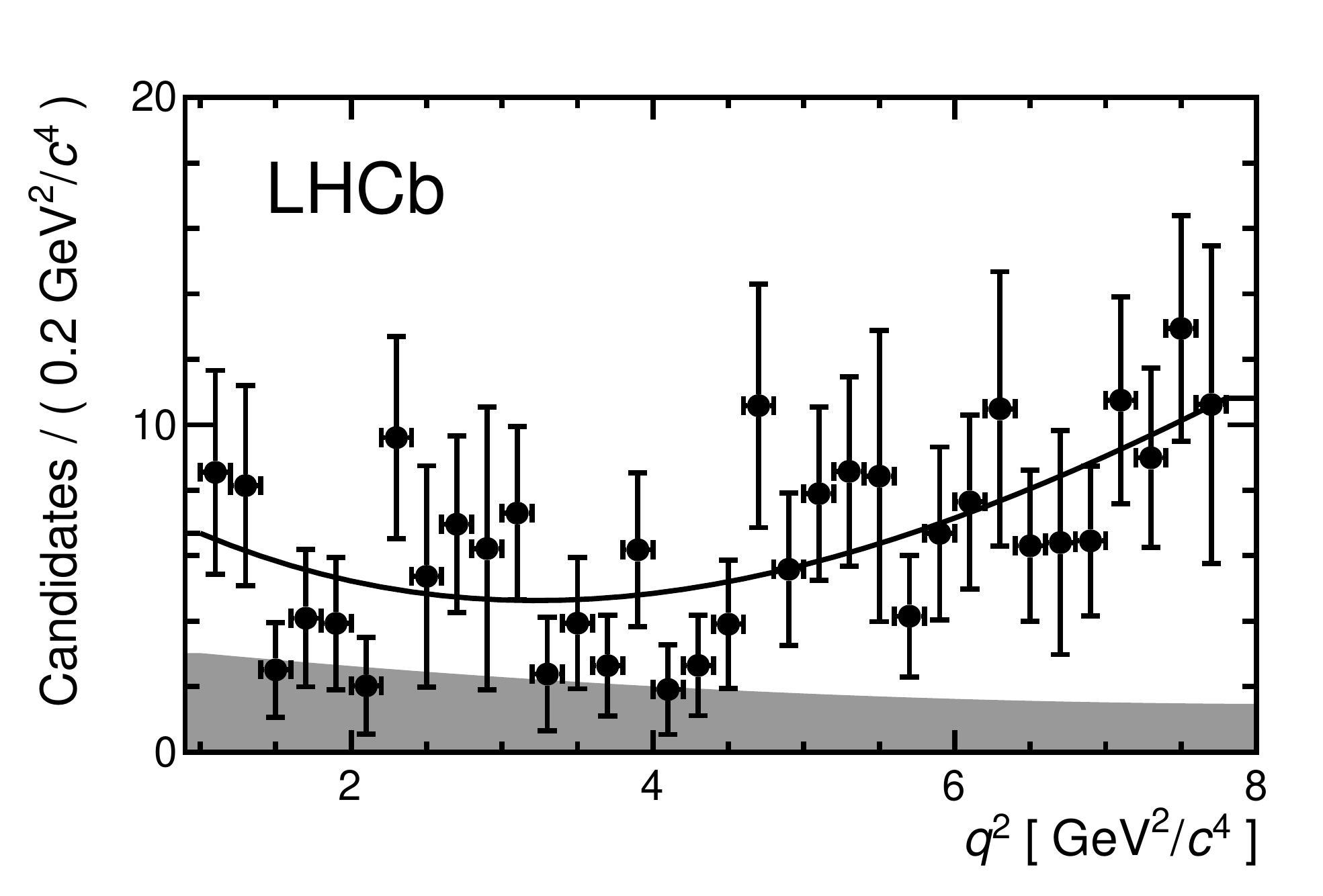}
\includegraphics[scale=0.38]{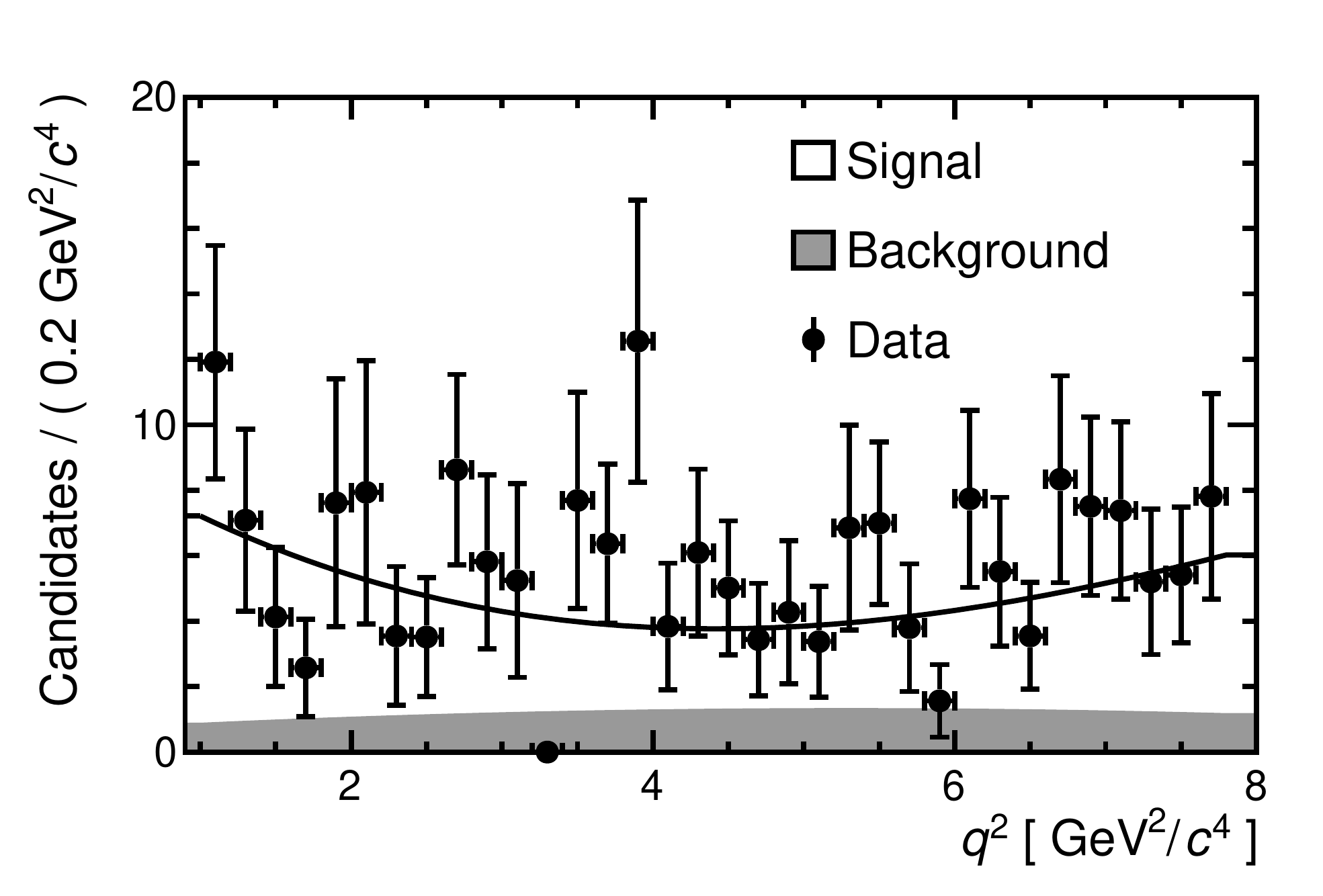} 
\caption{
Dimuon invariant mass squared, \qsq, distribution of forward-going (left) and backward-going (right) candidates in the $\Kp\pim\mumu$ invariant mass window $5230 < m(\Kp\pim\mumu) < 5330\mevcc$. The polynomial fit to the signal and background distributions in \qsq is overlaid.
\label{fig:zcp:data}
}
\end{figure}

The \qsq distribution of the forward- and backward-going candidates, in the range $1.0 < \qsq < 7.8 \gev^{2}/c^{4}$, is shown in Fig.~\ref{fig:zcp:data}. To make a precise measurement of the zero-crossing point a polynomial fit, $P(\qsq)$, is made to the \qsq distributions of these candidates. The $\Kp\pim\mumu$ invariant mass is included in the fit to separate signal from background. If $P_{\rm F}(\qsq)$ describes the \qsq dependence of the forward-going, and $P_{\rm B}(\qsq)$ the backward-going signal decays, then

\begin{equation}
A_{\rm FB}(\qsq) = \frac{P_{\rm F}(\qsq) - P_{\rm B}(\qsq)}{P_{\rm F}(\qsq) + P_{\rm B}(\qsq)} ~~. 
\end{equation}

\noindent The zero-crossing point of $A_{\rm FB}$ is found by solving for the value of \qsq at which $A_{\rm FB}(\qsq)$ is zero. 

Using third-order polynomials to describe both the \qsq dependence of the signal and the background, the zero-crossing point is found to be 

\begin{displaymath}
q_{0}^{2} = 4.9\pm 0.9 \gev^{2}/c^{4} ~~.
\end{displaymath}

\noindent  The uncertainty on $q^{2}_{0}$ is determined using a bootstrapping technique~\cite{Efron:1979}. The zero-crossing point is largely independent of the polynomial order and the \qsq range that is used. This value is consistent with SM predictions, which are typically in the range $3.9 - 4.4 \gev^{2}/c^{4}$~\cite{Bobeth:2011nj,Beneke:2004dp,Ali:2006ew} and have relative uncertainties below the 10\% level, for example, $q_{0}^{2} = 4.36\,^{+0.33}_{-0.31} \gev^{2}/c^{4}$~\cite{Beneke:2004dp}.

The systematic uncertainty on the zero-crossing point of the forward-backward asymmetry is negligible compared to the statistical uncertainty. To generate a large systematic bias, it would be necessary to create an asymmetric acceptance effect in $\cos\theta_{\ell}$ that is not canceled when combining \Bz and \Bzb decays. The combined systematic uncertainty is at the level of $\pm 0.05\gev^{2}/c^{4}$.

\section{Conclusions}
\label{sec:conclusions}

In summary, using a data sample corresponding to 1.0\invfb of integrated luminosity, collected by the \lhcb experiment in 2011, the differential branching fraction, $\deriv\BF/\deriv\qsq$, of the decay \decay{\Bz}{\Kstarz\mumu} has been measured in bins of \qsq. Measurements of the angular observables, $A_{\rm FB}$ ($A_{\rm T}^{\rm Re}$), $F_{L}$, $S_{3}$ ($A_{\rm T}^{2}$) and $A_{9}$ have also been performed in the same \qsq bins. 

The complete set of results obtained in this paper are provided in Tables~\ref{tab:diffbr} and \ref{tab:angular}. These are the most precise measurements of \deriv\BF/\deriv\qsq and the angular observables to date. All of the observables are consistent with SM expectations and together put stringent constraints on the contributions from new particles to $b\to s$ flavour changing neutral current processes. A bin-by-bin comparison of the reduced angular distribution with the SM hypothesis indicates an excellent agreement with p-values between 18 and 72\%. 

Finally, a first measurement of the zero-crossing point of the forward-backward asymmetry has also been performed, yielding $q_{0}^{2} = 4.9 \pm 0.9 \gev^{2}/c^{4}$. This measurement is again consistent with SM expectations.

\section*{Acknowledgements}

\noindent We express our gratitude to our colleagues in the CERN
accelerator departments for the excellent performance of the LHC. We
thank the technical and administrative staff at the LHCb
institutes. We acknowledge support from CERN and from the national
agencies: CAPES, CNPq, FAPERJ and FINEP (Brazil); NSFC (China);
CNRS/IN2P3 and Region Auvergne (France); BMBF, DFG, HGF and MPG
(Germany); SFI (Ireland); INFN (Italy); FOM and NWO (The Netherlands);
SCSR (Poland); MEN/IFA (Romania); MinES, Rosatom, RFBR and NRC
``Kurchatov Institute'' (Russia); MinECo, XuntaGal and GENCAT (Spain);
SNSF and SER (Switzerland); NAS Ukraine (Ukraine); STFC (United
Kingdom); NSF (USA). We also acknowledge the support received from the
ERC under FP7. The Tier1 computing centres are supported by IN2P3
(France), KIT and BMBF (Germany), INFN (Italy), NWO and SURF (The
Netherlands), PIC (Spain), GridPP (United Kingdom). We are thankful
for the computing resources put at our disposal by Yandex LLC
(Russia), as well as to the communities behind the multiple open
source software packages that we depend on.

\clearpage

{\noindent\bf\Large Appendix}

\appendix

\section{Angular basis} 
\label{sec:appendix:basis}

\begin{figure}[!hbt]
\centering
\includegraphics[width=\linewidth]{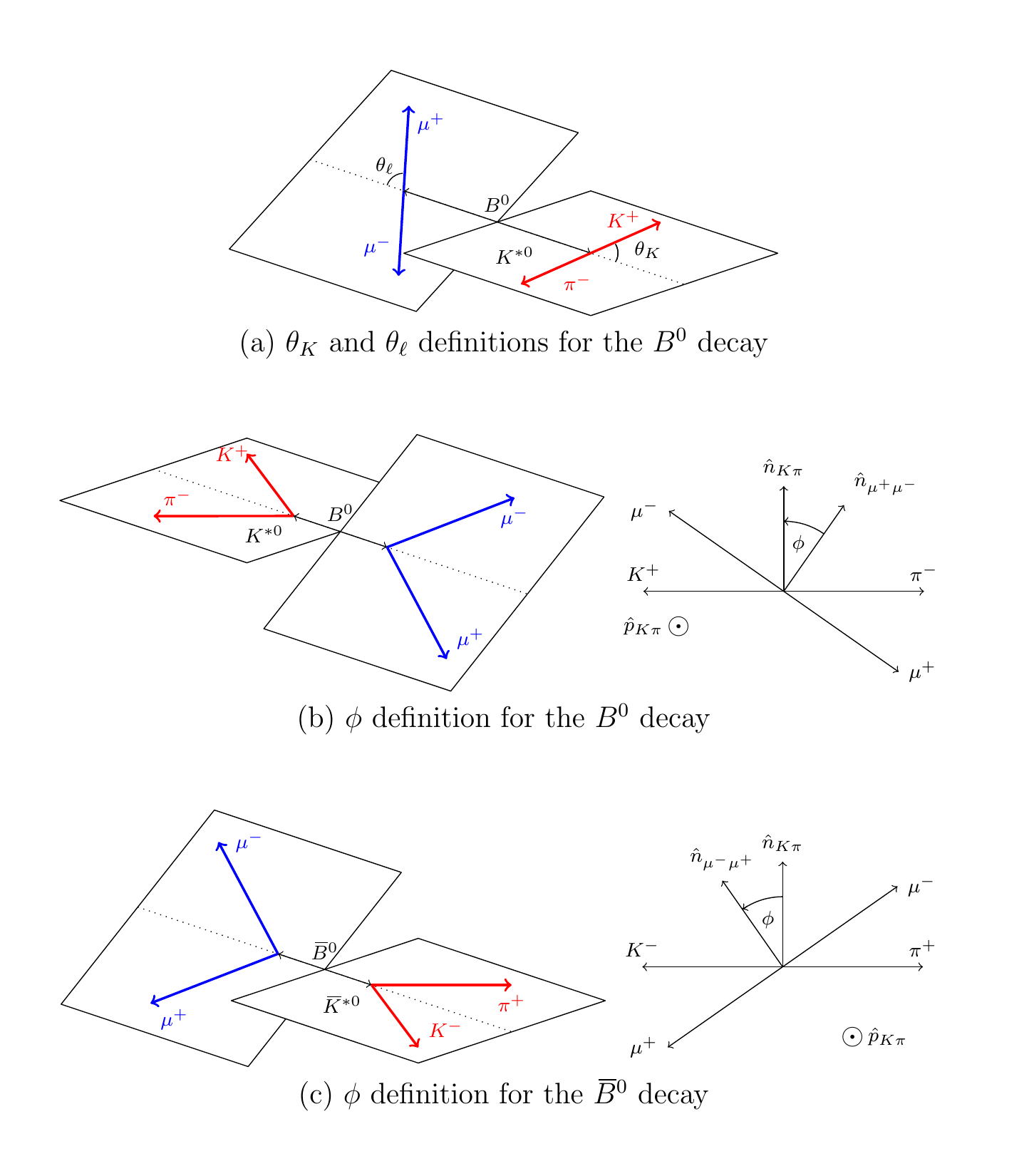}
\caption{
{\small
Graphical representation of the angular basis used for \decay{\Bz}{\Kstarz\mumu} and \mbox{\decay{\Bzb}{\Kstarzb\mumu}} decays in this paper. The notation $\hat{n}_{ab}$ is used to represent the normal to the plane containing particles $a$ and $b$ in the $\Bz$ (or \Bzb) rest frame. An explicit description of the angular basis is given in the text. 
}
\label{fig:appendix:basis}
}
\end{figure}

The angular basis used in this paper is illustrated in Fig.~\ref{fig:appendix:basis}. The angle $\theta_{\ell}$ is defined as the angle between the direction of the \mup (\mun) in the dimuon rest frame and the direction of the dimuon in the \Bz (\Bzb) rest frame. The angle $\theta_{K}$ is defined as the angle between the direction of the kaon in the \Kstarz (\Kstarzb) rest frame and the direction of the \Kstarz (\Kstarzb) in the \Bz (\Bzb) rest frame. The angle $\phi$ is the angle between the plane containing the \mup and \mun and the plane containing the kaon and pion from the \Kstarz. Explicitly, $\cos\theta_{\ell}$ and $\cos\theta_{K}$ are defined as  

\begin{equation}
\cos \theta_{\ell} = \left( \hat{p}_{\mup}^{(\mumu)} \right) \cdot \left( \hat{p}_{\mumu}^{(\Bz)} \right) = \left( \hat{p}_{\mup}^{(\mumu)}\right) \cdot \left( - \hat{p}_{\Bz}^{(\mumu)}\right) ~,
\end{equation}

\begin{equation}
\cos \theta_{K} = \left( \hat{p}_{\Kp}^{(\Kstarz)} \right) \cdot \left( \hat{p}_{\Kstarz}^{(\Bz)} \right)  = \left( \hat{p}_{\Kp}^{(\Kstarz)} \right) \cdot \left( - \hat{p}_{\Bz}^{(\Kstarz)} \right) 
\end{equation} 

\noindent for the \Bz and  

\begin{equation}
\cos \theta_{\ell} = \left( \hat{p}_{\mun}^{(\mumu)} \right) \cdot \left( \hat{p}_{\mumu}^{(\Bzb)} \right) = \left( \hat{p}_{\mun}^{(\mumu)} \right) \cdot \left( - \hat{p}_{\Bzb}^{(\mumu)}\right) ~,
\end{equation}

\begin{equation}
\cos \theta_{K} = \left( \hat{p}_{\Km}^{(\Kstarz)} \right) \cdot \left( \hat{p}_{\Kstarz}^{(\Bzb)} \right) = \left( \hat{p}_{\Km}^{(\Kstarz)} \right) \cdot \left( - \hat{p}_{\Bzb}^{(\Kstarz)} \right)
\end{equation} 

\noindent for the \Bzb decay. The definition of the angle $\phi$ is given by

\begin{equation}
\cos\phi =\left( \hat{p}_{\mup}^{(\Bz)} \times \hat{p}_{\mun}^{(\Bz)} \right) \cdot \left(\hat{p}_{\Kp}^{(\Bz)} \times \hat{p}_{\pim}^{(\Bz)} \right)  ~,
\end{equation}

\begin{equation} 
\sin \phi = \left[ \left(\hat{p}_{\mup}^{(\Bz)} \times \hat{p}_{\mun}^{(\Bz)}\right) \times \left(\hat{p}_{\Kp}^{(\Bz)}\times \hat{p}_{\pim}^{(\Bz)}\right) \right] \cdot \hat{p}_{\Kstarz}^{(\Bz)}
\label{eq:phi:B}
\end{equation} 

\noindent for the \Bz and 

\begin{equation}
\cos\phi = \left(\hat{p}_{\mun}^{(\Bzb)} \times \hat{p}_{\mup}^{(\Bzb)}\right) \cdot \left(\hat{p}_{\Km}^{(\Bzb)}\times \hat{p}_{\pip}^{(\Bzb)}\right)  ~,
\end{equation}

\begin{equation} 
\sin \phi = -\left[ \left(\hat{p}_{\mun}^{(\Bzb)} \times \hat{p}_{\mup}^{(\Bzb)}\right) \times \left(\hat{p}_{\Km}^{(\Bzb)}\times \hat{p}_{\pip}^{(\Bzb)}\right) \right] \cdot \hat{p}_{\Kstarzb}^{(\Bzb)}
\label{eq:phi:Bbar}
\end{equation}

\noindent for the \Bzb decay. The $\hat{p}_{X}^{(Y)}$ are unit vectors describing the direction of a particle $X$ in the rest frame of the system $Y$. In every case the particle momenta are first boosted to the \Bz (or \Bzb) rest frame. In this basis, the angular definition for the \Bzb decay is a \CP transformation of that for the \Bz decay.

\section{Angular distribution at large recoil}
\label{sec:appendix:largerecoil}

An explicit example of the bias on the angular observables that comes from the threshold terms is provided below for $A_{\rm T}^{2}$.  Sensitivity to $A_{\rm T}^{2}$ comes through the term in Eq.~\ref{eq:fullangular} with $\sin^{2}\theta_{\ell}\sin^{2}\theta_{K}\cos 2\phi$ angular dependence. In the limit $\qsq \gg m_{\mu}^{2}$, this term is simply

\begin{equation}
\frac{1}{2}\left( 1-F_{\rm L}(\qsq) \right) A_{\rm T}^{2}(\qsq) \sin^{2}\theta_{\ell}\sin^{2}\theta_{K}\cos 2\phi
\end{equation}

\noindent and the differential decay width is 

\begin{equation}
\frac{\deriv\Gamma}{\deriv\qsq} = |A_{0, {\rm L}}|^{2} + |A_{\parallel, {\rm L}}|^{2} + |A_{\perp, {\rm L}}|^{2} + |A_{0, {\rm R}}|^{2} + |A_{\parallel, {\rm R}}|^{2} + |A_{\perp, {\rm R}}|^{2} ~,
\end{equation}

\noindent where $A_{0}$, $A_{\parallel}$ and $A_{\perp}$ are the \Kstarz spin-amplitudes and the L/R index refers to the chirality of the lepton current (see for example Ref.~\cite{Altmannshofer:2008dz}). If $\qsq \lsim 1\gev^{2}/c^{4}$, these expressions are modified to 

\begin{equation}
\frac{1}{2} \left[ \frac{1 - 4m_{\mu}^{2}/\qsq}{1 + 2m_{\mu}^{2}/\qsq}\right] \left( 1-F_{\rm L}(\qsq) \right) A_{\rm T}^{2}(\qsq) \sin^{2}\theta_{\ell}\sin^{2}\theta_{K}\cos 2\phi
\end{equation}

\noindent and 

\begin{equation}
\frac{\deriv\Gamma}{\deriv\qsq} = \left[ 1 + 2m_{\mu}^{2}/\qsq \right] \left( |A_{0, {\rm L}}|^{2} + |A_{\parallel, {\rm L}}|^{2} + |A_{\perp, {\rm L}}|^{2} + |A_{0, {\rm R}}|^{2} + |A_{\parallel, {\rm R}}|^{2} + |A_{\perp, {\rm R}}|^{2} \right).
\end{equation}

In an infinitesimal window of \qsq, the difference between an experimental measurement of $A_{\rm T}^{2}$, $A_{\rm T}^{2~\text{exp}}$, in which the threshold terms are neglected and the value of $A_{\rm T}^{2}$ defined in literature is 

\begin{equation}
\label{eq:appendix:at2}
\frac{A_{\rm T}^{2~\text{exp}}}{A_{\rm T}^{2}} = \left[ \frac{1 - 4m_{\mu}^{2}/\qsq}{1 + 2m_{\mu}^{2}/\qsq} \right] ~~.
\end{equation}

\noindent Unfortunately, in a wider \qsq window, the \qsq dependence of $F_{\rm L}$, $A_{\rm T}^{2}$ and the threshold terms needs to be considered and it becomes less straightforward to estimate the bias due to the threshold terms. If $A_{\rm T}^{2}$ is constant over the \qsq window, 

\begin{equation}
\label{eq:appendix:at2:bin}
\frac{A_{\rm T}^{2~\text{exp}}}{A_{\rm T}^{2}} = \frac{\displaystyle\int_{q^2_{\text{min}}}^{q^2_{\text{max}}} \frac{\deriv\Gamma}{\deriv\qsq} \left[ \frac{1 - 4m_{\mu}^{2}/\qsq }{ 1 + 2 m_{\mu}^{2}/\qsq } \right] \left[ 1-F_{\rm L}(\qsq) \right] \deriv\qsq}{\displaystyle\int_{q^2_{\text{min}}}^{q^2_{\text{max}}} \frac{\deriv\Gamma}{\deriv\qsq} \left[ 1-F_{\rm L}(\qsq) \right] \deriv\qsq} ~~.
\end{equation}

\noindent In practice the integration in Eq.~\ref{eq:appendix:at2:bin} can be replaced by a sum over the signal events in the \qsq window

\begin{equation}
\label{eq:at2dilution}
\frac{A_{\rm T}^{2~\text{exp}}}{A_{\rm T}^{2}} = \frac{\sum\limits_{i=0}^{N} \left[ \frac{1 - 4m_{\mu}^{2}/q^{2}_{i}}{1 + 2 m_{\mu}^{2} / q^{2}_{i}} \right] (1 - F_{\rm L} (q_{i}^{2}))\omega_{i}}{\sum\limits_{i=0}^{N} (1 - F_{\rm L} (q_{i}^{2}))\omega_{i}} ~,
\end{equation}

\noindent where $\omega_{i}$ is a weight applied to the $i^{\text{th}}$ candidate to account for the detector and selection acceptance and the background in the \qsq window.

Correction factors for the other observables can be similarly defined if it is assumed that they are constant over the \qsq window. In the case of $A_{\rm FB}$ (and $A_{\rm T}^{\rm Re}$) that are expected to exhibit a strong \qsq dependence, the \qsq dependence of the observable needs to be considered.

\addcontentsline{toc}{section}{References}
\bibliographystyle{LHCb}
\bibliography{main,rare,stat,detector}

\end{document}